\documentclass[12pt]{article}
\pdfoutput=1
\usepackage{color,amsmath,amssymb,exscale,psfrag,epsfig}
\usepackage{cite,color,url}
\usepackage{pdfpages}
\usepackage{graphicx}
\usepackage{slashed}
\usepackage[utf8]{inputenc}
\usepackage[T1]{fontenc}
\usepackage{xcolor}
\usepackage{accents}
\usepackage{centernot}
\usepackage{fixltx2e}
\usepackage{enumerate}
\usepackage{subfig}
\usepackage{float}

\newcommand{\met}{\ensuremath{E_{\text{T}}^{\text{miss}}}}

\usepackage[colorlinks=true
,urlcolor=blue
,anchorcolor=blue
,citecolor=blue
,filecolor=blue
,linkcolor=blue
,menucolor=blue
 ,pagecolor=blue
,linktocpage=true
,pdfproducer=medialab
,pdfa=true
]{hyperref}
\def\bea{\begin{eqnarray}}
\def\eea{\end{eqnarray}}

\definecolor{nicered}{rgb}{0.7,0.1,0.1}
\definecolor{nicegreen}{rgb}{0.1,0.5,0.1}
\hypersetup{colorlinks,citecolor= nicegreen,linkcolor= nicered}

\newcommand{\change}[1]{#1}

\catcode`\@=11
\def\lsim{\mathrel{\mathpalette\@versim<}}
\def\gsim{\mathrel{\mathpalette\@versim>}}
\def\@versim#1#2{\vcenter{\offinterlineskip
\ialign{$\m@th#1\hfil##\hfil$\crcr#2\crcr\sim\crcr } }}
\catcode`\@=12
\catcode`@=12
\begin{document}
\thispagestyle{empty}
\begin{flushright}
ICAS 036/18
\end{flushright}
\vspace{0.1in}
\begin{center}
{\Large \bf Non-resonant Leptoquark with \\ multigeneration couplings for $\mu\mu jj$ and $\mu\nu jj$ at LHC} \\
\vspace{0.2in}
{\bf Ezequiel Alvarez$^{\dagger}$,
Manuel Szewc$^{\diamond}$
}
\vspace{0.2in} \\
{\sl International Center for Advanced Studies (ICAS)\\
 UNSAM, Campus Miguelete, 25 de Mayo y Francia, (1650) Buenos Aires, Argentina }
\\[1ex]
\end{center}
\vspace{0.1in}

\begin{abstract}
	CMS has recently reported \cite{Sirunyan:2018ryt} a moderate excess in the $\mu\nu jj$ final state in a second generation Leptoquark search, but they have disregarded it because the excess is not present in the $\mu\mu jj$ final state and because they do not observe the expected resonant peak in the distributions. \change{As a proof of concept} we show that a simple Leptoquark model including second and third generation couplings with non-negligible single- and non-resonant production in addition to usual pair production could explain the data: excess ($\mu\nu jj$), lack of excess ($\mu\mu jj$) and missing peak in the distributions; while being in agreement with collider constraints. We take this result and analysis as a starting point of a reconsideration of the ATLAS and CMS second generation Leptoquark searches. We also discuss which would be the consequences and modifications that should be performed in the searches to test if this deviation would correspond to a New Physics signal. \change{We observe that low-energy flavor constraints can be avoided by adding heavier particles to the model.}
\end{abstract}

\vspace*{2mm}
\noindent {\footnotesize E-mail:
{\tt 
$\dagger$ \href{mailto:sequi@unsam.edu.ar}{sequi@unsam.edu.ar},
$\diamond$ \href{mailto:mszewc@unsam.edu.ar}{mszewc@unsam.edu.ar},
}}

\section{Introduction}
\label{section:1}
Since the Higgs Boson discovery in 2012\cite{Aad:2012tfa, Chatrchyan:2012xdj, Khachatryan:2016vau} practically all experimental high energy physics results have stuck to the Standard Model (SM) predictions. Among the most notorious exceptions are the B-anomalies 
$$R_{K^{(*)}} = \frac{{\cal B} (B \to K^{(*)} \mu\mu)}{{\cal B} (B \to K^{(*)} ee) } $$ 
at a significance level of $\sim 3.5-4\sigma$ \cite{Aaij:2014ora, Aaij:2017vbb, Hiller:2017bzc, Capdevila:2017bsm} and
$$R_{D^{(*)}} = \left. \frac{{\cal B} (B \to D^{(*)} \tau\bar\nu)}{{\cal B} (B \to D^{(*)} \ell \ell) } \right|_{\ell =e,\mu} $$
at $\sim$4$\sigma$ \cite{Aaij:2015yra, Huschle:2015rga,Freytsis:2015qca, Amhis:2016xyh}. After a broad variety of New Physics (NP) models \cite{Greljo:2015mma, Sierra:2015fma, Li:2018rax,Kawamura:2017ecz, Buttazzo:2017ixm} attempting to explain these anomalies while not being ruled out by all other experimental results, Leptoquarks seem to be the explanation that best accommodates data\cite{Becirevic:2016yqi, Hiller:2017bzc, Buttazzo:2017ixm}.

Although $B$-physics results by Babar\cite{Lees:2012xj, Lees:2013uzd}, Belle\cite{Huschle:2015rga} and LHCb\cite{Aaij:2014ora, Aaij:2015yra, Aaij:2017vbb} indicate a possible presence of Leptoquark NP, this is all at the level of a low energy Effective field Theory (EFT)\cite{Buttazzo:2017ixm, Hiller:2017bzc, Calibbi:2015kma, Becirevic:2015asa, Dorsner:2017ufx, Crivellin:2017zlb, Choudhury:2017qyt, Becirevic:2016yqi}. A current challenge is to look for Leptoquark traces in observables where these new particles could be actually either produced on-shell and then decayed or either exchanged in a t-channel Feynman diagram. This kind of observables should be investigated in the general purpose LHC experiments ATLAS and CMS.

In the last few years ATLAS and CMS have developed an interesting Leptoquark search program. This program consist in specific searches for Leptoquarks of first\cite{Aad:2011uv, Aad:2015caa, Khachatryan:2015vaa, Sirunyan:2018btu}, second\cite{Aad:2011uv, Aad:2015caa, Khachatryan:2015vaa, Aaboud:2016qeg, Sirunyan:2018ryt} and third generation\cite{Sirunyan:2018nkj, CMS-PAS-B2G-16-027, Sirunyan:2017yrk} as well as other related searches\cite{Aaboud:2017opj, Aaboud:2017wqg, Sirunyan:2017kqq}. Since Leptoquarks carry color charge they are produced through QCD gauge couplings, and most analyses of a given generation consist in searches of pair produced Leptoquarks with the corresponding decay of each Leptoquark to a lepton and a quark in the corresponding generation. The sensitivity to each generation is complex and depends on the sensitivity and efficiency for tagging each lepton and quark. 

Along this article we discuss the available second generation Leptoquark searches by ATLAS and CMS which study the final states $\mu\mu jj$ and $\mu \nu jj$. Since the early LHC, both experiments have published searches of this kind. In particular ATLAS begun with searches in both final states for early Run I\cite{Aad:2011uv} but for some unknown reason the following results considered only the $\mu\mu jj$ final state \cite{Aad:2015caa, Aaboud:2016qeg}, overlooking $\mu\nu jj$. On the other hand, CMS has always published results considering both final states \cite{Khachatryan:2015vaa, Sirunyan:2018ryt}. Interestingly, if no $b$-tag veto is applied on the jets, since the neutrino flavor cannot be identified, then the $\mu\nu jj$ final state is also sensitive to one of the Leptoquarks decaying to third generation. Therefore, both final states have different contributions from second and third generation couplings. 

In this work we discuss the latest second generation Leptoquark search at LHC and we refer to it as the CMS paper Ref.~\cite{Sirunyan:2018ryt}. This paper analyzes second generation Leptoquarks in the final states $\mu\mu jj$ and $\mu \nu jj$ for a luminosity of 35.9 fb$^{-1}$ at a center of mass energy of 13 TeV for the first time. They find a $2.25\sigma$ excess in the $\mu \nu jj$ final state and no excess in $\mu\mu jj$. In addition they report that the kinematic distributions do not show the expected peak from on-shell Leptoquarks. Although this hint is still too small to be considered as a NP signal, these results motivated us to explore the possibility that the excess could be originated by a Leptoquark with \change{diagonal} couplings to second and also third generation, and whose couplings are large enough to have single- and non-resonant amplitudes that could hide pair production resonant effects. \change{We understand as pair production when there are two Leptoquarks on-shell, single production when there is only one Leptoquark on-shell, and non-resonant to all other cases, as for instance a Leptoquark in t-channel. Non-resonant effects and multigeneration Leptoquarks have already been considered in the literature in other scenarios \cite{Diaz:2017lit,Raj:2016aky,Schmaltz:2018nls}.}  In this article we propose a simple Leptoquark model with these features and we study its capability to qualitatively reproduce the results in CMS paper Ref.~\cite{Sirunyan:2018ryt}. We obtain some quantitative results, however a more detailed quantitative analysis should be performed by the CMS collaboration using all their available data. We find that the series of results obtained in this work should encourage ATLAS and CMS to reconsider second generation Leptoquark searches and include third generation physics in their hypothesis as well as in their analysis of the final states.

This paper is divided as follows. In Section \ref{section:2} we present a simple Leptoquark model that could account for the observed features. In Section \ref{section:3} we study its qualitative phenomenology: we find the latest bounds in its parameter space, we chose a benchmark point to study and we describe qualitatively how this model will provide more events to the $\mu\nu jj$ final state than to $\mu\mu jj$. In Section \ref{section:4} we perform a series of simulations and a quantitative comparison to the available data. In Section \ref{section:5} we discuss some topics related to potential changes in the simulation and in the model and we also briefly discuss other anomaly observed in CMS paper \cite{Sirunyan:2018ryt}. We end with our conclusions in Section \ref{section:6}.  \change{We include an appendix where we discuss some low-energy flavor constraints and how they can be avoided by adding more particles to the model.}

\section{Simple Leptoquark model}
\label{section:2}
For simplicity, we assume that there is only one kind of Leptoquark involved in the production of the $\mu\mu jj$ and $\mu\nu jj$ final states.  \change{If a given Leptoquark decays to $\mu j$ and also $\nu j$ then }its electric charge can be either $1/3$ or $2/3$.  As we show along the text (sections \ref{pheno} and \ref{section:5}), the former is more favored. Therefore we assume a Leptoquark with $Q=1/3$.

The main objective in this work is to address the results in CMS paper Ref.~\cite{Sirunyan:2018ryt}. Just for the sake of concreteness, we will work with a Leptoquark that is also involved in a possible solution for the B-anomalies, as for instance $S_{3}$ or $S_{1}$~\cite{Hiller:2017bzc, Fajfer:2017lzq, Marzocca:2018wcf, Angelescu:2018tyl, Becirevic:2018afm, Crivellin:2017zlb, Dorsner:2017ufx, Buttazzo:2017ixm}. We choose to work with the $S_1$ Leptoquark with quantum numbers $S_1 \sim (\bar 3, 1)_{1/3}$, where each number denotes the representation under the SM Group $(SU(3)_C,SU(2)_L)_{U(1)_Y}$. Being a $SU(2)$ singlet, $S_1$ does not present constraints coming from other multiplet partners of different charge.  Its quark-lepton interaction Lagrangian can be written as\cite{Dorsner:2016wpm}
\begin{equation}
{\cal L} \supset y^{LL}_{ij}\bar Q_L^{C\,i,a}\, S_{1} \, \epsilon^{ab}\, \ell^{j,b}_L + y^{RR}_{ij}\bar u_R^{C\,i}\, S_{1} \, e^{j}_R + y^{\overline{RR}}_{ij}\bar d_R^{C\,i}\, S_{1} \,  \nu^{j}_R.
\label{lagrangian}
\end{equation}
Where $\epsilon$ is the anti-symmetric tensor and $y^{XX}_{ij}$ the NP couplings constants.

Throughout this work we take the following ansatz for the coupling constants
\begin{eqnarray}
y^{LL} &=& \left( \begin{array}{ccc}
0 & 0 & 0 \\
0 & y_{22} & y_{23} \\
0 & y_{32} & y_{33} \end{array} \right) 
\label{eqy}
\end{eqnarray}
and $y^{RR}_{ij}=y^{\overline{RR}}_{ij}=0$. 
\change{This ansatz is specified at the TeV scale.  Although we present this model from a phenomenological point of view in order to perform a proof of concept on the CMS paper Ref.~\cite{Sirunyan:2018ryt}, one may observe that the pattern of a Leptoquark coupled mainly to second and third generations at the TeV scale can be expected in models where Leptoquarks are pseudo-Nambu Goldstone Bosons of a strongly interacting heavy sector \cite{Alvarez:2018gxs,DaRold:2018moy}.}
For simplicity, we take all $y_{ij}$ to be real and non-negative. However, they could have different signs without affecting our results.
The ansatz in Eq.~\ref{eqy} avoids first generation constraints by forbidding Leptoquark interactions with electrons and certain flavor constraints by forbidding right-handed currents~\cite{Dorsner:2016wpm, Wise:2014oea}. As shown in sections \ref{section:3} and \ref{section:4}, to account for the reported results in \cite{Sirunyan:2018ryt} Leptoquark single- and non-resonant production have to be competitive with pair production. For TeV scale masses, this can be achieved by assuming some components in $y_{ij}$ are of ${\cal O}(1)$\cite{Dorsner:2018ynv, Buttazzo:2017ixm}. For simplicity we will assume in the following that
\begin{equation}
y_{23},\, y_{32} \ll 1.
\end{equation}
\change{Since along the text we have only $y_{22,33}$ non negligible, then when referring to a Leptoquark coupling to second or third generation, it means diagonally coupled to both lepton and quark of the given generation.}  The case where $y_{32}$ and $y_{23}$ are non negligible is discussed in section \ref{section:5}. 

With these assumptions and the Lagrangian in Eq.~\ref{lagrangian}, this particle has the following open channels:
\begin{equation}
S^{-1/3} \to c\,\mu^{-},\ t\,\tau^{-},\ s\,\nu_i,\ b\,\nu_i .
\label{channels}
\end{equation}
\change{The Leptoquark width is determined by
\begin{eqnarray}
\Gamma_{S^{1/3}}&=& \sum_{q_{i} = u_{i},d_{i}}\sum_{L_{j}=\ell_{j}, \nu_{j}}\frac{|y_{q_{i}L_{j}}|^2}{16\pi M^{3}_{S^{1/3}}}(M^{2}_{S^{1/3}}-m^{2}_{q_{i}}-m^{2}_{L_{j}})\nonumber\\
&& \sqrt{M^{4}_{S^{1/3}}-2M^{2}_{S^{1/3}}(m^{2}_{q_{i}}+m^{2}_{L_{j}})+(m^{2}_{q_{i}}-m^{2}_{L_{j}})^{2}}\nonumber\\
|y_{u_{i}\ell_{j}}|&=&(V^{T}y^{LL*})_{ij}\nonumber\\
|y_{d_{i}\nu_{j}}|&=&(y^{LL}U)_{ij}\nonumber\\
\end{eqnarray} 
}
Assuming V\textsubscript{CKM} $ \approx 1$, using the unitarity of the neutrino mixing matrix U\textsubscript{PMNS}, and assuming a top quark mass negligible compared to $M_{S^{1/3}}$, we can easily derive the Branching Ratios
\begin{equation}
\begin{array}{rclcl}
BR(S^{-1/3} \to c \mu^{-} ) &=& \sum_{i} BR(S^{-1/3} \to s \nu_{i} ) &=& \frac{y_{22}^2}{2(y_{22}^2 + y_{33}^2) } \\
BR(S^{-1/3} \to t \tau^{-} ) &=& \sum_{i} BR(S^{-1/3} \to b \nu_{i} ) &=& \frac{y_{33}^2}{2(y_{22}^2 + y_{33}^2) } 
\end{array}
\label{decays}
\end{equation}
If one is agnostic to flavor and assumes $y_{22} \approx y_{33}$, the Branching Ratios obey that
\begin{equation}
\sum_{i}BR(j\nu_{i}) \approx 2\, BR(j\mu)
\end{equation}

The difference between neutrinos and muons resides in that neutrinos are produced with both second and third generation jets due to $y_{33}\neq 0$ while muons are produced only with second generation c-jets. As seen in section \ref{section:3}, this difference between neutrinos and muons can help understand the differences between the $\mu\mu jj$ and $\mu\nu jj$ channels in CMS paper Ref.~\cite{Sirunyan:2018ryt}. 

To study this interplay between the two generations relevant to $\nu j$ decays, it is useful to define the parameter 
\begin{equation}
r=\frac{y_{33}}{y_{22}}.
\label{r}
\end{equation}
The Branching Ratios can then be written as 
\begin{equation}
\begin{array}{rclcl}
BR(S^{-1/3} \to c \mu^{-} ) &=& \sum_{i} BR(S^{-1/3} \to s \nu_{i} ) &=& \frac{1}{2(1 + r^2) } \\
BR(S^{-1/3} \to t \tau^{-} ) &=& \sum_{i} BR(S^{-1/3} \to b \nu_{i} ) &=& \frac{r^2}{2(1 + r^2) } 
\end{array}
\label{decays2}
\end{equation}
 
\begin{figure}[h!]
\begin{center}
\includegraphics[width=0.45\textwidth]{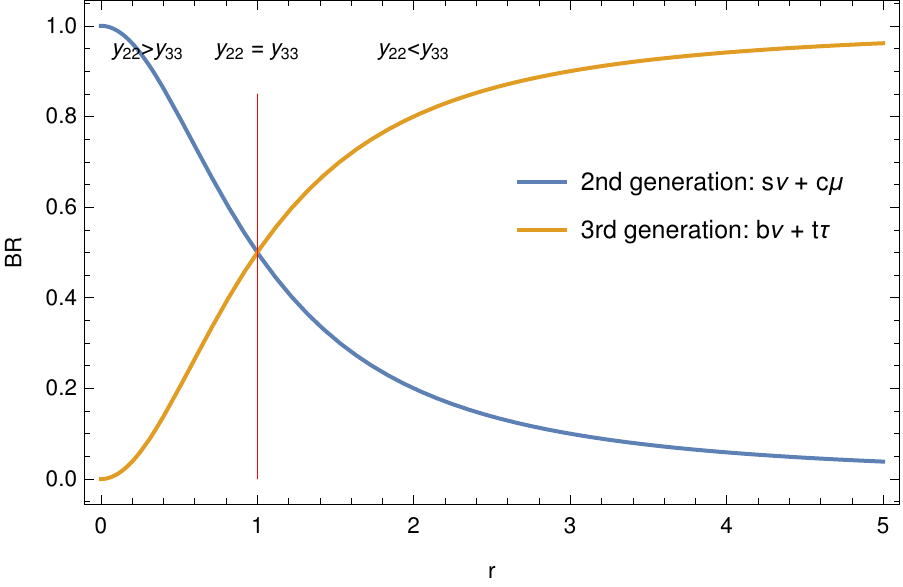}
\caption{\small Branching ratio for the second and third generations as a function of the parameter $r$ (see Eq.~\ref{r}). The Branching Ratios to each specific channel is the half of the plotted Branching Ratio to the corresponding generation.}
\label{fig:branchingratios}
\end{center}
\end{figure}

The Branching Ratio to second and third generation are plotted as a function of $r$ in Fig.~\ref{fig:branchingratios}. For $r < 1$, the second generation is preferred while for $r > 1$ the third generation dominates. 

As discussed in section \ref{section:3}, when the third generation Branching Ratio dominates, the model can avoid tensions in purely second generation final states such as $\mu \mu jj$ without sacrificing a larger event yield in channels sensitive to third generation couplings such as $\mu\nu jj$. This corresponds to the parameter region where $y_{22} \leq y_{33}$ which implies $r \geq 1 $.

\section{Phenomenology for the $S^{1/3}$ Leptoquark}
\label{section:3}
\begin{figure}[h!]
\begin{center}
\includegraphics[width=0.45\textwidth]{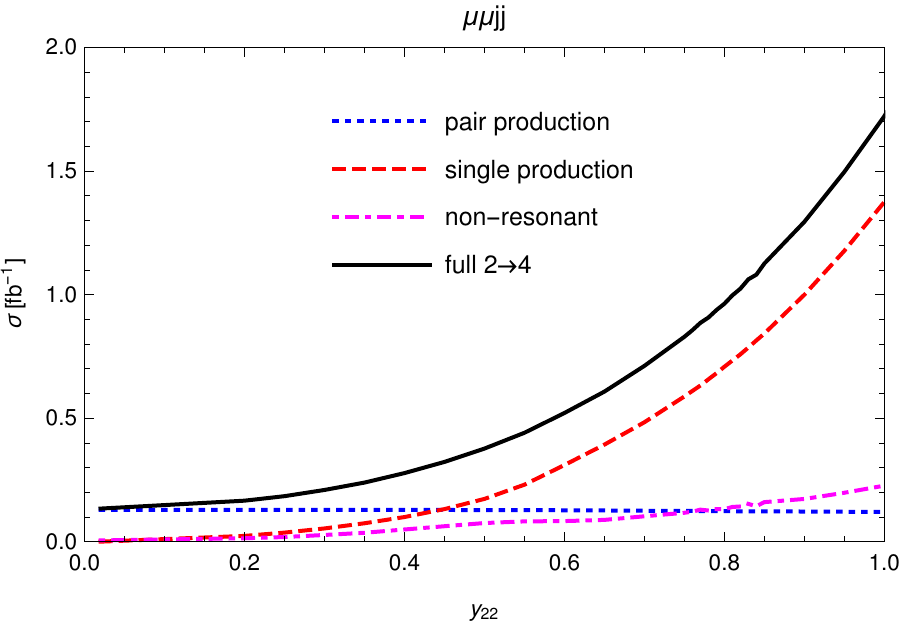} ~
\includegraphics[width=0.45\textwidth]{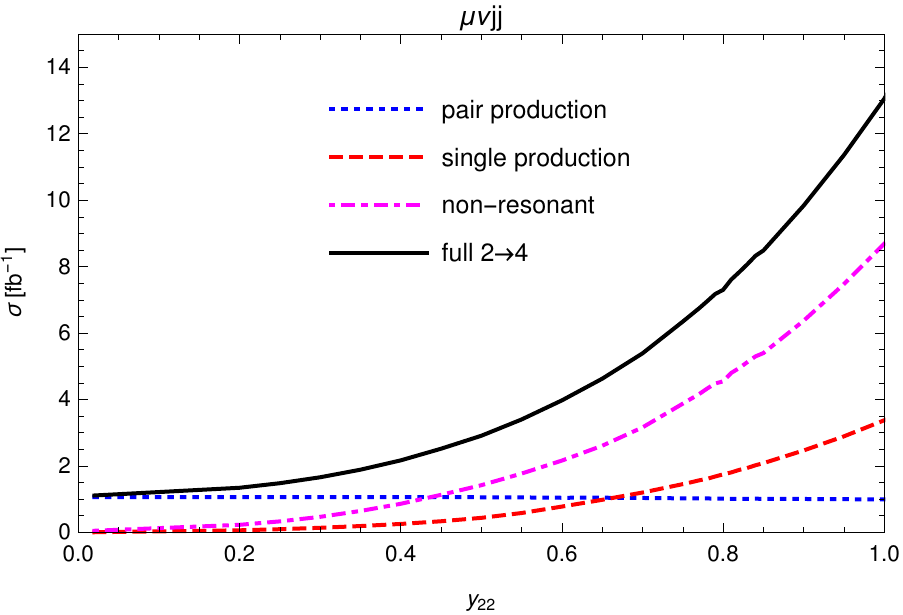}
        \caption{\small \change{Pair, single- and non-resonant production contributions to the final states $\mu\mu jj$ and $\mu\nu jj$ as a function of $y_{22}$.  Other parameters are fixed at $r=1.7$ and $M_{S^{1/3}}=950$ GeV. In both cases for $y_{22} \gtrsim 0.4$ resonant pair production becomes sub-leading.  Due to the relative size of the couplings for each generation, $y_{33}/y_{22} = r = 1.7$, the non-resonant contribution is the dominant for $\mu\nu jj$, whereas single-production contribution dominates for $\mu\mu jj$. } }
\label{contributions}
\end{center}
\end{figure}

We are interested in the final states $\mu\nu jj$ and $\mu\mu jj$ at the LHC running at a center of mass energy of $\sqrt{s}=13$ TeV. To study the phenomenology of the $S^{1/3}$ Leptoquark it is important to determine the region to explore in parameter space: the smaller the Leptoquark mass $M_{S^{1/3}}$ and the couplings $y_{ij}$, the more favored is pair production whereas the larger they are, the more favored is single- and non-resonant production\cite{Dorsner:2018ynv, Buttazzo:2017ixm}. Since CMS paper Ref.~\cite{Sirunyan:2018ryt} observes that kinematic distributions in the deviation lack the characteristic mass peak expected for on-shell Leptoquarks, it is convenient to explore a region in parameter space where $\mu\nu jj$ and $\mu\mu jj$ non-resonant production could hide on-shell characteristic kinematic distributions. We find that the region determined by
\begin{equation}
\begin{array}{rcl}
M_{S^{1/3}} & \sim & {\cal O}(1\mbox{ TeV}) \\
y_{22},\, y_{33} & \sim & {\cal O}(1)
\end{array}
\label{region}
\end{equation}
successfully achieves the above requirements, while keeping the physics perturbative and within the scope of current LHC analyses.  \change{We plot in Fig.~\ref{contributions} the size of the different contributions to the total cross-section (with basic cuts as outlined in Ref.~\cite{Sirunyan:2018ryt}) of both $\mu\mu jj$ and $\mu\nu jj$ final states as a function of the coupling $y_{22}$ for fixed $r$ and LQ mass.  We find that for $y_{22} \gtrsim 0.4$ resonant pair-production becomes sub-leading.}   In the following paragraphs we investigate which are the current limits and potential issues in parameter space, we define a benchmark point, and we study the main production mechanisms for the relevant states $\mu\nu jj$ and $\mu\mu jj$ in the region of parameter space indicated in Eq.~\ref{region} and in which lies the benchmark point.

\subsection{Current direct limits}
The region in parameter space defined in Eq.~\ref{region} may yield single-, non-resonant and pair production within the same order of magnitude. However, given the available LHC experimental results in direct Leptoquark searches, the main direct bounds come from pair production like mechanisms. In the following paragraphs we obtain the constraints coming from this kind of searches. However, one should be cautious in the limit setting procedure as the competitiveness of single- and non-resonant production would affect signal and background control regions needed to set the limits. \change{Since we are considering a region in parameter space where single and non-resonant dominates over resonant pair production, we expect that NP affects more the background estimation than the signal counting.  Therefore,} assuming only on-shell pair production NP in a scenario where single- and non-resonant mechanisms may not be negligible would yield stronger limits than what they actually are\change{, since backgrounds estimated through data-driven techniques would be under-estimated}.  Work in the direction of considering non-resonant effects in existing searches can be found in Ref.\cite{Mandal:2018kau, Bansal:2018eha}.

The decaying channels and Branching Ratios for $S^{-1/3}$ are indicated in Eqs.~\ref{channels} and \ref{decays2}. We can identify here as potentially problematic the channels $t\tau^{-}$, $b+\met$, $s+\met$ and $c\mu^{-}$. We investigate these four channels bearing in mind that the results on the latter are the ones we are reconsidering in this work. 

We can derive constraints on our model by recasting existing searches. As the searches focus on pair production, they assume a Branching Ratio scheme to set an upper bound on the cross-section. $t\tau^{-}$ and $b+\met$ are recasted from third generation Leptoquark searches in the $t\bar{t}\tau^+\tau^-$ final state \cite{Sirunyan:2018nkj} and the supersymmetric search of sbottom pair production through the $b\bar b+\met$ final state \cite{Aaboud:2017wqg,Sirunyan:2017kqq} for the case of zero neutralino mass, respectively. \change{Limits on $s+\met$ final state are obtained from recasting the squark pair production search designed for first and second generation in final state $jj+\met$ \cite{Aaboud:2017vwy}.  To use the limits found in these searches, the Branching Ratio to their corresponding final states has to be adapted to the point in parameter space of our model.} $c\mu^{-}$ is recasted from second generation Leptoquark searches \cite{Sirunyan:2018ryt, Aaboud:2016qeg} which assume a $BR(S^{-1/3}\to c \mu^{-})$ of either 1 or $1/2$. Due to the multiplet structure, $S^{1/3}$ has a maximum Branching Ratio of $1/2$ to any decay channel. This weakens the limits set on $M_{S^{1/3}}$ for third generation searches as the maximum cross-section allowed can increase by a factor of at most 4. 

\begin{figure}[h!]
\begin{center}
\includegraphics[width=0.45\textwidth]{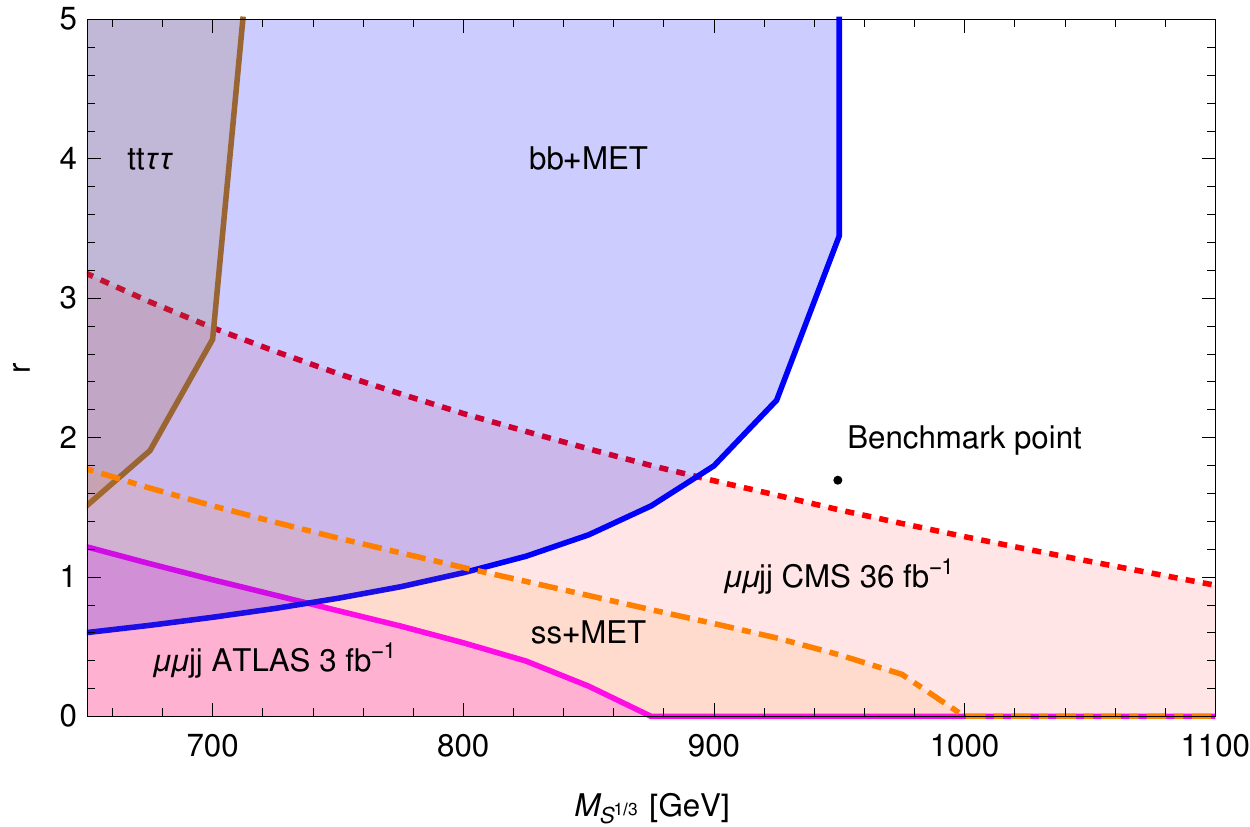}
	\caption{\small Excluded regions in the $M_{S^{1/3}}$ vs.~$r$ plane. The dashed line represents the limits from CMS paper Ref.~\cite{Sirunyan:2018ryt} assuming no NP effects. The $\mu\mu jj$ ATLAS limit is recasted from Ref.~\cite{Aaboud:2016qeg}, the $tt\tau\tau$ limits is recasted from Ref.~\cite{Sirunyan:2018nkj}, the $bb+\met$ limit is recasted from Ref.~\cite{Sirunyan:2017kqq}, and the $ss+\met$ limits is recasted from Ref.~\cite{Aaboud:2017vwy}.}
\label{fig:rvsm}
\end{center}
\end{figure}

If we parameterize the Branching Ratios with $r$ (Eqs.~\ref{r} and \ref{decays2}) we can recast the limits set by each search to a limit on $M_{S^{1/3}}$ for each $r$ as can be seen in Fig.~\ref{fig:rvsm}. As the third generation Branching Ratios increase with $r$, third generation searches set upper bounds on the allowed $r$. On the other hand, second generation Branching Ratios decrease with $r$ and second generation searches set a lower bound on $r$ for each $M_{S^{1/3}}$. Observe that since any Branching Ratio is bounded by $1/2$, then all Leptoquark masses whose cross-section is less than four times the limit set by the searches are allowed for any value of $r$.

Fig.~\ref{fig:rvsm} summarizes the above described analysis recasted as limits on the allowed values of $r$ for each possible $M_{S^{1/3}}$. We indicate a benchmark point defined in the following paragraphs. As expected from the previous discussion, for $M_{S^{1/3}}$ large enough $r$ is no longer bounded.

\subsection{Benchmark point}

Following the excluded regions in Fig.~\ref{fig:rvsm} and motivated by the roughly two standard deviation excess in CMS paper Ref.~\cite{Sirunyan:2018ryt}, we consider the benchmark point defined by:
\change{\begin{eqnarray}
\label{eq:950}
y_{33} &=& 1.2 \nonumber \\
y_{22} &=& 0.7 \\ \nonumber
M_{S^{1/3}} &=& 950 \mbox{ GeV}.
\end{eqnarray}}
This point is indicated in Fig.~\ref{fig:rvsm}.

The benchmark point yields \change{a width of $\Gamma \sim 7.5\%$ of the mass}, with Branching Ratios within the constraints on pair production from Refs.~\cite{Aaboud:2017wqg,Sirunyan:2017kqq}, $BR(S^{-1/3} \to t\tau^{-})=0.37$ and $BR(S^{-1/3} \to c\mu^{-})=0.13$. It should be noted that a large $y_{33}$ allows the third generation to act as a escape valve against constraints from second generation Leptoquark searches and to populate the $\mu\nu jj$ final state.

It is worth noting that this benchmark point, having large couplings to enhance non-resonant production, is likely to yield potential problems in flavor physics \cite{Dorsner:2016wpm,Angelescu:2018tyl, Buttazzo:2017ixm, Becirevic:2016oho,Becirevic:2018afm, Marzocca:2018wcf, Dorsner:2017ufx, Crivellin:2017zlb} which should be addressed in detail.  \change{With this purpose, we perform in the Appendix an overview of the major flavor constraints on this kind of proposals.  In particular, we find that adding heavier Leptoquarks to the model provides a possible solution to tensions in different low-energy observables.  Another possibility is to reduce the couplings absolute value while keeping their ratio.  This still provides an excess in $\mu\nu jj$ over $\mu\mu jj$, but at the price of loosing non-resonant features in the kinematic distributions.}

\subsection{$S^{1/3}$ contribution to final states $\mu\mu jj$ and $\mu\nu jj$ at the LHC}
\label{pheno}

Having stated the region in parameter space to be explored, the current bounds on it, and a specific benchmark point, we study the production mechanisms for the final states $\mu\mu jj$ and $\mu\nu jj$ in which $S^{1/3}$ will have a relevant contribution. One of the main goals in this section is to qualitatively understand how the presented model and benchmark point can yield larger contributions to $\mu\nu jj$ than to $\mu\mu jj$ to better agree with the results in CMS paper Ref.~\cite{Sirunyan:2018ryt}. We can distinguish two relevant differences: {\it i)} Leptoquark Branching Ratios (relevant to pair production and decay), and {\it ii)} quark abundances in the proton PDF (relevant to single- and non-resonant production). Observe that to study the different relative contribution to the final states $\mu\mu jj$ and $\mu\nu jj$, only the latter has a dependence on the chosen $M_{S^{1/3}}$ and absolute value of the couplings.

\begin{figure}[h!]
\begin{center}
\subfloat[]{\includegraphics[width=0.45\textwidth]{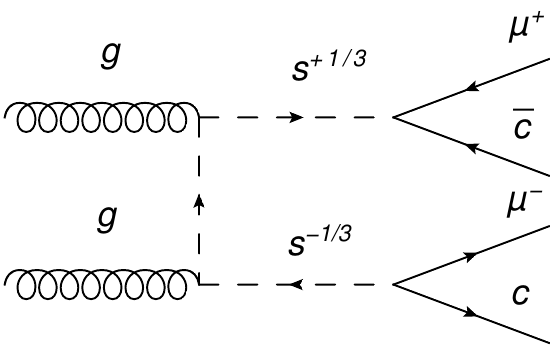}}\hspace{3mm}
\subfloat[]{\includegraphics[width=0.45\textwidth]{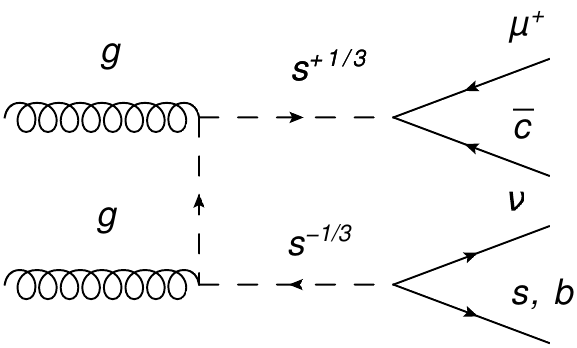}}\hspace{3mm}\\
\caption{\small Representative QCD pair production mechanism diagrams for the final states a) $\mu\mu jj$ and b) $\mu\nu jj$. As $r$ increases the $\mu\nu jj$ final state is considerably preferred over $\mu\mu jj$. }
\label{pair}
\end{center}
\end{figure}

\begin{figure}[h!]
\begin{center}
\subfloat[]{\includegraphics[width=0.45\textwidth]{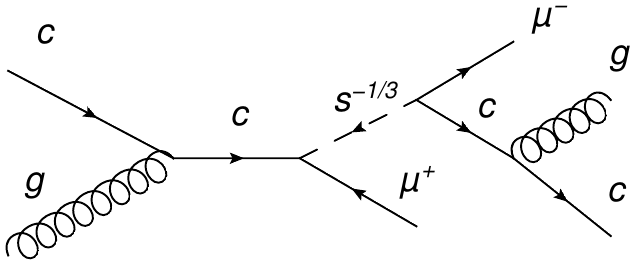}}\\
\subfloat[]{\includegraphics[width=0.45\textwidth]{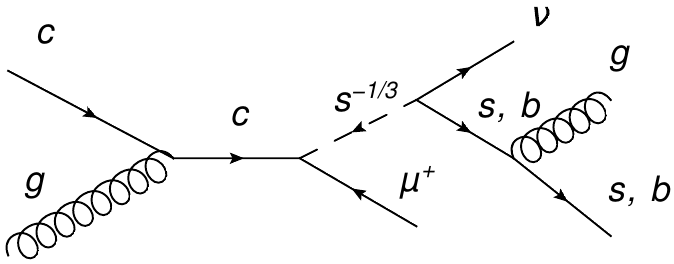}}\hspace{3mm} 
\subfloat[]{\includegraphics[width=0.45\textwidth]{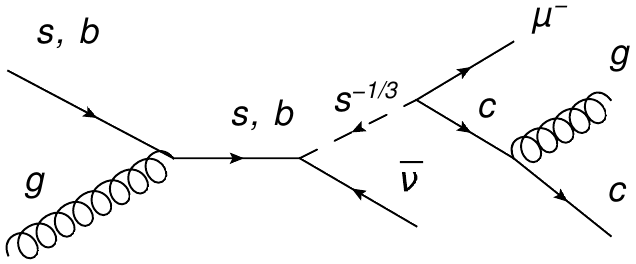}}\\
\caption{\small Representative single-production mechanism diagrams for the final states a) $\mu\mu jj$ and b) and c) $\mu\nu jj$. Notice how $c\, g$ initial state contributes to $\mu\mu jj$ and $\mu\nu jj$, whereas $b/s\, g$ initial state only contributes to $\mu\nu jj$. }
\label{single}
\end{center}
\end{figure}

\begin{figure}[h!]
\begin{center}
\subfloat[]{\includegraphics[width=0.25\textwidth]{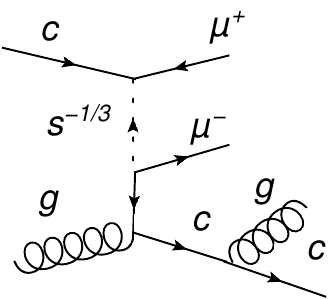}}\\
\subfloat[]{\includegraphics[width=0.25\textwidth]{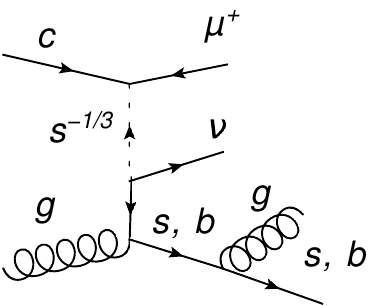}}\hspace{3mm} 
\subfloat[]{\includegraphics[width=0.25\textwidth]{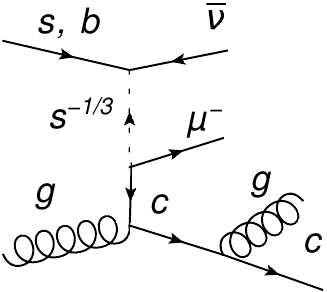}}\\
\caption{\small Representative non-resonant production mechanism diagrams for the final states a) $\mu\mu jj$ and b) and c) $\mu\nu jj$. As in single-resonant production, notice that $c\, g$ initial state contributes to $\mu\mu jj$ and $\mu\nu jj$, whereas $b/s\, g$ initial state only contributes to $\mu\nu jj$. }
\label{nonresonant}
\end{center}
\end{figure}

When producing the final states $\mu\mu jj$ and $\mu\nu jj$ through QCD pair production $pp \to S^{1/3} S^{-1/3}$ (see Fig.~\ref{pair}) the process is dominated by gluon fusion due to the QCD gauge coupling between gluons and Leptoquarks. Cross-section for processes involving pair production can be approximated as
\begin{eqnarray}
\sigma(gg\to (S^{1/3}\to l_{1}q_{1})(S^{-1/3}\to l_{2}q_{2}))&\simeq& \sigma(gg\to S^{1/3}S^{-1/3})BR(S^{1/3}\to l_{1}q_{1})\nonumber\\&&BR(S^{-1/3}\to l_{2}q_{2}).
\nonumber
\end{eqnarray}
As such, the relevant information from the couplings is their ratio $r$ and only the Branching Ratios yield the difference between the $\mu\mu jj$ and $\mu\nu jj$ channels.

As we study $y_{22} \approx y_{33}$ we see in Eq.~\ref{decays} that we can expect approximately four times more $\mu\nu jj$ than $\mu\mu jj$. The reason is that $y_{33}\neq 0$ opens a channel to third generation decays. This effect alone could explain the observed difference between channels, but, as it is reported in \cite{Sirunyan:2018ryt}, the absence of a peak in the distributions indicates that single- and non-resonant production should also be taken into account. In pair production all final states come from on-shell particles, whereas this is not valid for single- and non-resonant production. In these channels this argument exclusively on the Branching Ratios no longer applies and one should consider different arguments.

In contrast to pair production which is produced from a $gg$ initial state, single- and non-resonant productions require quarks in the initial state. This generates a new imbalance that favors $\mu\nu jj$ over $\mu\mu j j$. We show in Figs.~\ref{single} and \ref{nonresonant} some representative Feynman diagrams for the single- and non-resonant production mechanism. As it can be seen, there is a difference between the partonic processes relevant in each channel. We refer in the following paragraphs to partonic processes with disregard to whether the quarks and leptons involved are particles or antiparticles. There is an imbalance when comparing 

\begin{equation}
s\, g \to \mu \nu j j \qquad \mbox{vs.} \qquad c\, g \to \mu \mu j j
\end{equation}
because the PDF abundance of the $s$-quark is larger than the $c$-quark. Moreover, when analyzing the analogous diagrams but at the third generation level in a 5-flavor scheme the process

\begin{equation}
b\, g \to \mu \nu j j
\end{equation}
does not have its counterpart which would require top-quark in the proton PDF. This imbalance between up-type and down-type quarks enhances $\mu\nu jj$ over $\mu\mu jj$.

There are also other sources of imbalance between the $\mu\mu jj$ and $\mu\nu jj$ channels. As for instance, in single- and non-resonant production it can be found that the initial state $c\, g$ contributes to the final state $\mu\nu jj$ channel whereas the $s/b \, g$ initial state does not have it corresponding diagram contributing to the final state $ \mu \mu j j$. This can be seen through the single-resonant representative Feynman diagrams in Fig.~\ref{single}. This effect is not due to quark abundances in the proton but to the definition of the channels. 
 
The above paragraphs indicate how we can expect a larger deviation in $\mu\nu jj$ than in $\mu\mu j j$ for many production processes. This difference increases still more as we increase $r=y_{33}/y_{22}$. For pair production because the difference between third and second generation Branching Ratios grows. And for single- and non-resonant production because b-associated diagrams have a larger coupling constant. Considering the benchmark point defined in Eq.~\ref{eq:950} there is still room to increase the difference between the $\mu\mu jj$ and $\mu\nu j j$ channels while keeping the model within the experimental bounds discussed in the above paragraphs.

\section{Contrasting $S^{1/3}$ Leptoquark to CMS results}
\label{section:4}

In previous sections we have defined a model and a region in its parameter space which we have argued to favor an excess in the final state $\mu\nu jj$ over $\mu\mu jj$. We have justified that this region in parameter space would also hide a peak in kinematic distributions because of single- and non-resonant processes involved in the production of the mentioned final states. Along this section we probe this model through a given benchmark point (Eq.~\ref{eq:950}) and test the validity of our reasoning. We begin with a brief description of the CMS paper Ref.~\cite{Sirunyan:2018ryt}, we then describe our simulations and then present the results, which correspond to compare our simulations with the CMS available results.
 
\subsection{CMS second generation Leptoquark search results}
The CMS collaboration reported in Ref.~\cite{Sirunyan:2018ryt} a search for second generation Leptoquark pair production in the $\mu\mu jj$ and $\mu\nu jj$ channels. To discriminate against SM backgrounds, three kinematic variables are selected for each channel: $S_{T}^{\mu\mu jj}$, $m_{\mu\mu}$ and $m^{min}_{\mu j}$ for $\mu\mu jj$ and $S_{T}^{\mu\nu jj}$, $m^{T}_{\mu\nu}$ and $m_{\mu j}$ for $\mu\nu jj$. $S_{T}$ is the transverse momentum scalar sum while $m_{ij}$ and $m^{T}_{ij}$ are the invariant mass and invariant transverse mass of the corresponding particles, respectively. The $\mu j$ pairs are selected by minimizing the (transverse) mass difference between the Leptoquark candidates for the $\mu\mu jj$ ($\mu\nu jj$) channels.

Different cuts in these kinematic variables define different signal regions. Each cut corresponds to a function of the mass of the Leptoquark for which the signal region is optimized. The Leptoquark mass for which the cuts are optimized is varied between $200$ GeV and $2000$ GeV. Since each cut is a monotonically growing function of the mass of the candidate Leptoquark, a given mass signal region is a subset of events of the lower mass signal regions. To define each signal region that is optimized for a Leptoquark of given mass $M_{S^{1/3}}$, Ref.~\cite{Sirunyan:2018ryt} uses the variable $M_{LQ}$. It should be stressed that $M_{LQ}$ is not a mass, but a signal region optimized for a Leptoquark of mass $M_{S^{1/3}} = M_{LQ}$. Further details on these signal regions can be found in CMS paper Ref.~\cite{Sirunyan:2018ryt}.

The collaboration reports event yields in each signal region and there is a qualitative different result for the observed versus expected number of events in each final state. There is a deficit in the $\mu\mu jj$ channel for the signal regions $M_{LQ} \sim 600\mbox{ GeV} - 800$ GeV, with a significance in a specific bin of about $2\sim 3$ standard deviations. On the other hand there is an excess in the $\mu\nu jj$ channel for $M_{LQ} \sim 900\mbox{ GeV} - 1100$ GeV.  \change{ This excess is maximum at $M_{LQ}=950$ GeV, where the expected background events is $16.9\pm1.0\pm1.7$ and the measured data is 30 events \cite{Sirunyan:2018ryt}.  Including a Poissonian uncertainty for the data and adding all uncertainties in quadrature yields a significance of 2.25 standard deviations.} Observe that because of the correlation between each signal region $M_{LQ}$ is not possible at this level to perform a multi-$M_{LQ}$ analysis for the deficit or the excess. In the following sections we focus on the excess in $\mu\nu jj$ rather than in the deficit in $\mu\mu jj$. In CMS paper Ref.~\cite{Sirunyan:2018ryt} it is stated that the given excess does not show a characteristic peak in the $m_{\mu j}$ distribution, which would be characteristic from Leptoquark pair production since final particles would come from on-shell NP particles.

Our goal is to reproduce the excess in the $\mu\nu jj$ final state without significantly altering $\mu\mu jj$. We discuss the deficit in $\mu\mu jj$ in Section \ref{section:5}.

\subsection{Simulation}

To reproduce an excess in one channel without getting in tension with the other, and to show that single- and non-resonant production wash out the peak in the $m_{\mu j}$ distribution, we focus on a qualitative analysis at the parton level for the benchmark point detailed in Eq.~\ref{eq:950}. Prospects for a more detailed analysis are discussed in Section \ref{section:5}.

We have implemented our model using {\tt Feynrules}~\cite{Alloul:2013bka} and loaded it into {\tt Madgraph\;5}~\cite{Alwall:2014hca}. We have simulated $pp\to \mu\mu j j$ and $pp\to \mu\nu jj$ including $S^{1/3}$ single-, non-resonant and pair production at $\sqrt{s}=13$~TeV. The signal events have been generated using a Leading Order matrix element and the MSTW2008 PDF set~\cite{Martin:2009iq}, with the renormalization and factorization scales set to $M_{S^{1/3}}$ \cite{Kramer:2004df, Mandal:2015lca} with the same cuts as in CMS paper Ref.~\cite{Sirunyan:2018ryt}. After generation, expected events in each signal region are obtained assuming an efficiency of 30$\%$.  This estimation comes from the reported acceptance times efficiencies in CMS paper Ref.~\cite{Sirunyan:2018ryt}, taking into account that detector acceptances for pair production are close to unity.  We discuss different efficiencies in Section \ref{section:5}. \change{Regarding NLO-QCD effects we have considered the available UFO model in Ref.~\cite{Dorsner:2018ynv}. However, at the current level of development this model sets a maximum of one NP vertex for NLO processes and therefore we cannot generate several of the required relevant diagrams.  Despite this, we have been able to compute pair-production and $q g \to S^{1/3} \ell$ NLO corrections, reproducing the results of Ref.~\cite{Dorsner:2018ynv} which yield $K \approx 1.56$ for pair-production and $K \approx 1.38/1.3/1.65$ for $b/c/s g \to S^{1/3} j \ell$, respectively. Therefore, and taking the approximate agreement of these $K$-factors into account, we estimate an overall $K$-factor of $K = 1.5$ that includes non-resonant effects.}

Simulations were computed in a 5-flavor scheme. However, we also performed the simulation in a 4-flavor scheme and verified that results remain qualitatively unchanged. Since we simulate up to parton level we do not need to perform a matching to avoid double counting. 

Further discussion about the simulations, its parameters and its approximations is given in section \ref{section:5}.

\subsection{Results}

\begin{figure}[h!]
\begin{center}
\subfloat[]{\includegraphics[width=0.45\textwidth]{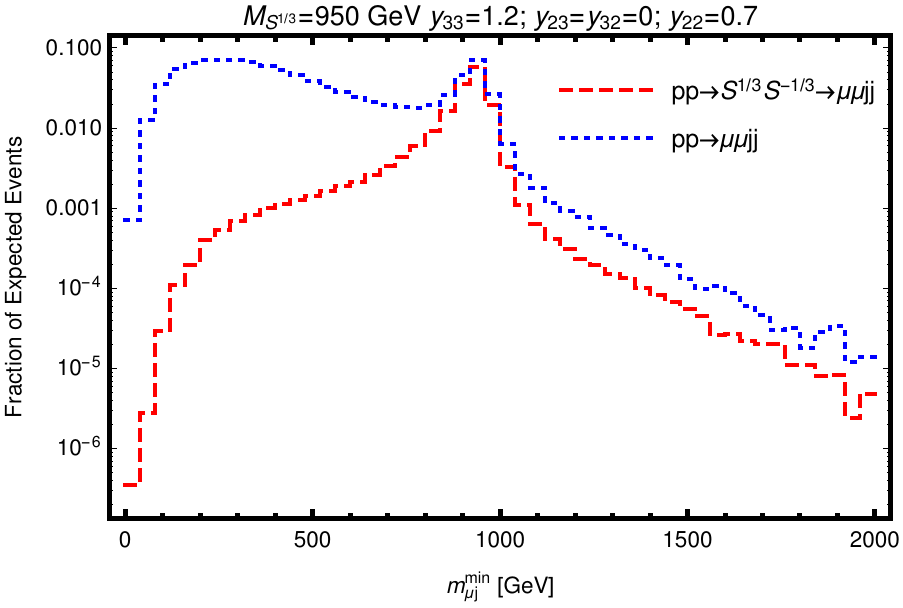}}\hspace{3mm}
\subfloat[]{\includegraphics[width=0.45\textwidth]{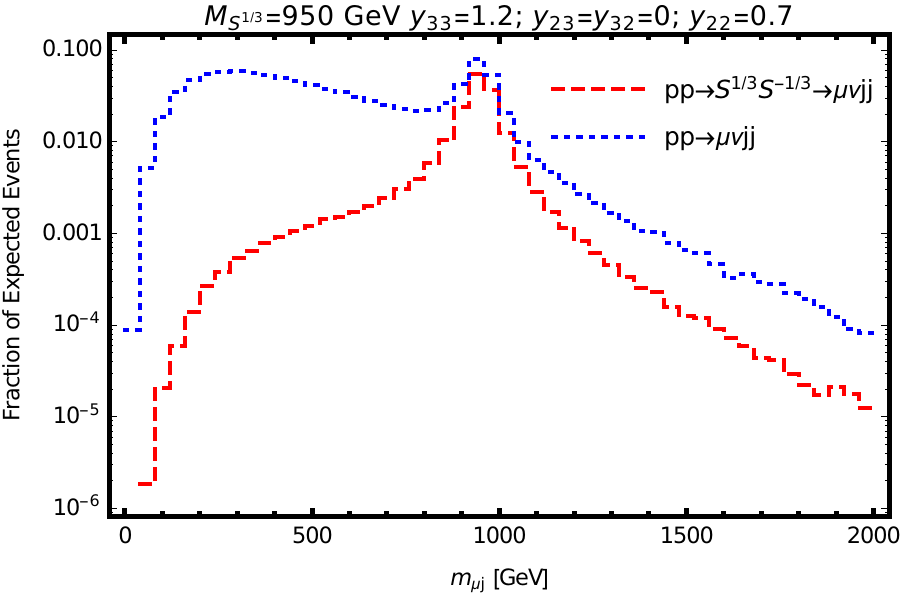}} \\
\caption{\small $m_{\mu j}$ kinematic distribution at the a) $\mu\mu jj$ and b) $\mu\nu jj$ final states for only NP diagrams. The red (dashed) line corresponds to only pair production diagrams, whereas the blue (dotted) line includes in addition single- and non-resonant production.}
\label{fig:kinematics}
\end{center}
\end{figure}

With the simulated samples one can study the difference between considering only pair production and taking into account both single- and non-resonant behavior. This difference is made clear in the kinematic distributions, such as the $m^{min}_{\mu j}$ ($m_{\mu j}$) for the $\mu\mu jj$ ($\mu\nu jj$) final state, as can be seen in Fig.~\ref{fig:kinematics}. It is seen in both final states that the pair production peak is washed-out by the single- and non-resonant effects. The single- and non-resonant Feynman diagrams discussed in Section \ref{pheno} provide the kinematics to yield events with $m^{min}_{\mu j}$ and $m_{\mu j}$ considerably different to the mass of the Leptoquark involved in the process.

In this benchmark point, the event yield in each signal region for each final state $\mu\mu jj$ and $\mu\nu jj$ can be compared to data reported in \cite{Sirunyan:2018ryt}. Because of the qualitative discussion in Section \ref{pheno}, and mainly because of the PDF abundance of the involved initial state quarks, we expect NP to provide a larger contribution to the $\mu\nu jj$ than to the $\mu\mu jj$ final state. 

\begin{figure}[h!]
\begin{center}
\subfloat[]{\includegraphics[width=0.45\textwidth]{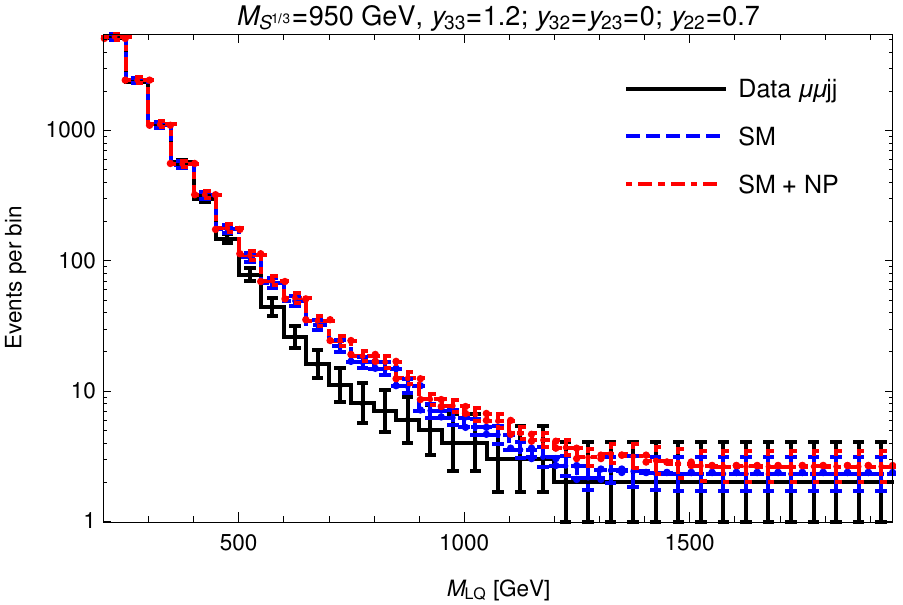}}\hspace{3mm}
\subfloat[]{\includegraphics[width=0.45\textwidth]{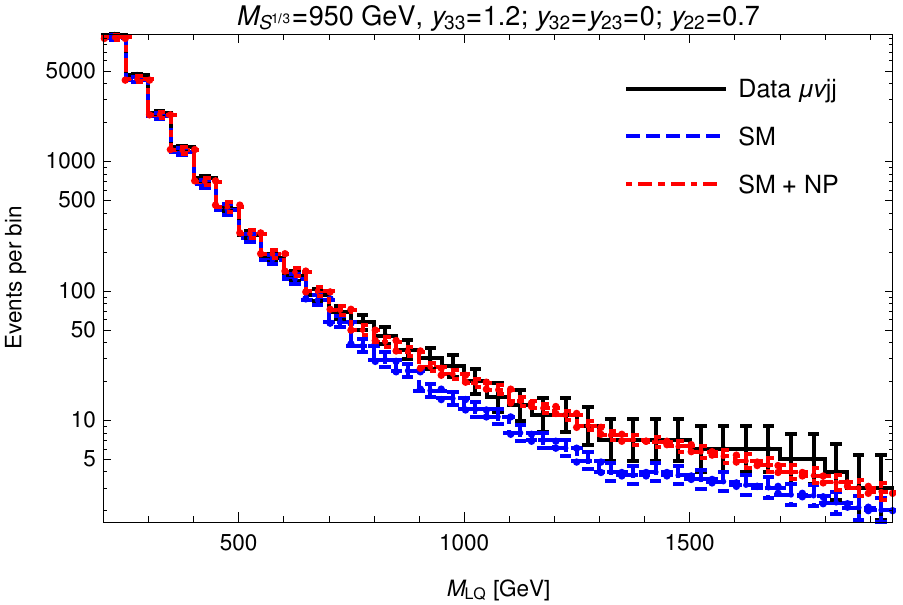}} \\
\subfloat[\label{fig:comp1}]{\includegraphics[width=0.45\textwidth]{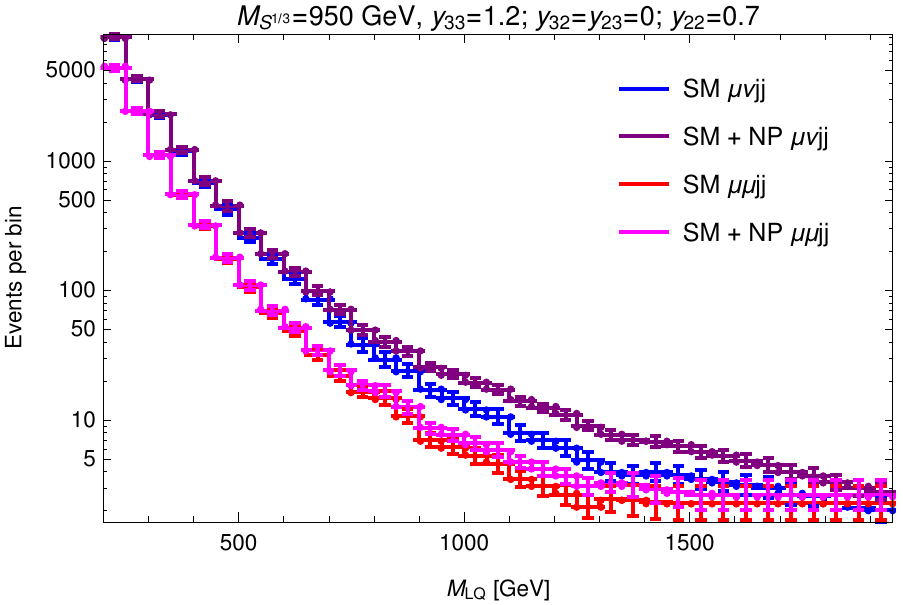}}\hspace{3mm}
\subfloat[\label{fig:comp2}]{\includegraphics[width=0.45\textwidth]{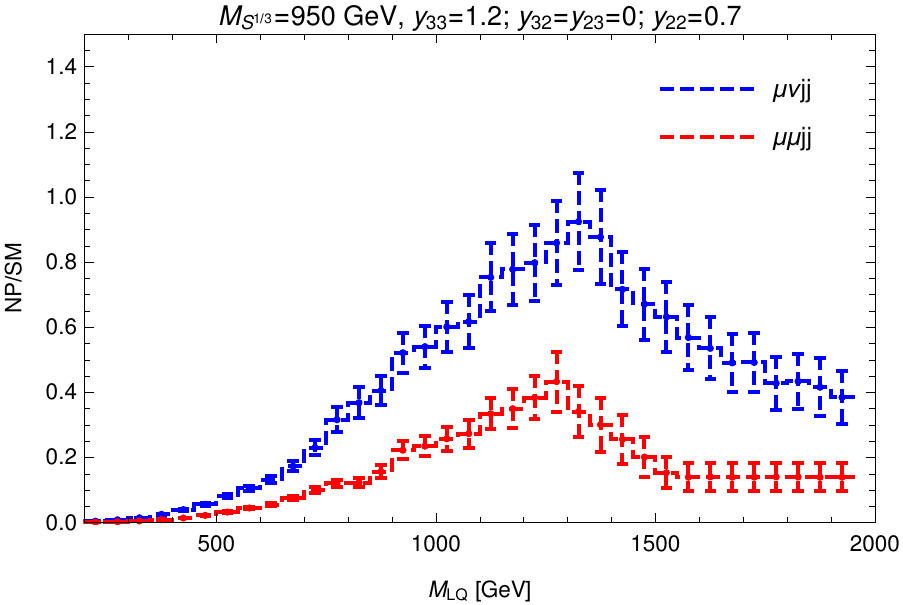}} \\
\caption{\small Top row: Comparison between data, SM and SM+NP events for a) $\mu\mu jj$ and b) $\mu\nu jj$ final states. Bottom row: Comparison between $\mu\mu jj$ and $\mu\nu jj$ NP event yields in a) absolute value b) relative deviation from SM}
\label{fig:twochannels}
\end{center}
\end{figure}

The SM backgrounds have been taken to be those reported in \cite{Sirunyan:2018ryt} and they have been combined with our simulated samples in order to have a comparison as seen in Fig.~\ref{fig:twochannels}. 
The top row in the figure shows both channels separated while the bottom row compares both channels together, both in absolute value and in its deviation for their respective SM background. As it can be seen, the $\mu\nu jj$ final state has both a larger absolute NP event yield (Fig.~\ref{fig:comp1}) and a larger relative deviation (Fig.~\ref{fig:comp2}) from background than $\mu\mu jj$.

The largest deviation in the $\mu\nu jj$ final state reported in \cite{Sirunyan:2018ryt}, located at the $M_{LQ}=950$ GeV bin, is reduced from $2.25 \sigma$ to $0.75 \sigma$ in this NP benchmark point. 

Therefore, in addition to the wash-out of the peak in the kinematic distributions, this shows quantitatively that the NP model can provide an explanation of the moderate excess in the $\mu\nu jj$ final state while keeping without considerable change the events in the $\mu\mu jj$ final state.

\section{Discussion}
\label{section:5}

The results obtained in the previous section reflect the differences qualitatively discussed in section \ref{section:3} between the two channels studied in second generation Leptoquark searches when the proposed NP has a multigeneration non-resonant phenomenology. This is the main goal in this work.  Along this section we briefly discuss related topics which, in case the experimental deviation is established, should be further developed.   We first discuss on the agreement between the model and the data: we begin by considering how our results could be affected if we perform a more detailed simulation, we then study how our model assumptions can be modified and what differences does this alteration inflict upon the analysis and results, we also discuss why the hypothesis of electric charge $Q=1/3$ is favored over a Leptoquark with $Q=2/3$, and finally we examine the reported deficit in the $\mu\mu jj$ final state and possible explanations in case that corresponds to a NP effect.
 
The results reported in the previous section have been obtained using the benchmark point on the parameter space defined in Eq.~\ref{eq:950}. \change{The simulations have been done at parton level using a fixed NLO $K$-factor with an assumed fixed efficiency of 30$\%$ (see Ref.~\cite{Sirunyan:2018ryt})}. A more sophisticated study should include NLO generation, showering, hadronization and detector simulation. The wanted features of the model are already present at the qualitative level and are maintained as long as $r\sim 1-3$, the $y_{22}\sim y_{33} \sim {\cal O}(1)$ and $M_{S^{1/3}}\approx 1$ TeV. Therefore, a reasonable modification in the efficiencies could be compensated by modifying the $y_{ij}$ coefficients and/or the mass within the aforementioned values to still hold the results in Section \ref{section:4}. In particular, observe that the region around the benchmark point in Fig.~\ref{fig:rvsm} allows for considerable more enhancement of $\mu\nu jj$ over $\mu\mu jj$.   To verify the above, we have also considered efficiencies ranging from $0.1$ to $0.4$, including cases with larger values for the $\mu\mu jj$ final state, and we have been able to qualitatively reproduce the same results as with the benchmark point by varying only $y_{22}$ and $y_{33}$ from $0.5$ to $2.0$.  In any case, if the studied excess would become more significant, it would be essential that experimentalists unfold the data and publish the information required for a full quantitative comparison.

The chosen benchmark point includes the simplistic assumptions made in section \ref{section:2}. If one seeks a flavor hierarchical model, this may require non negligible $y_{32}$ and $y_{23}$. If $V_{CKM} \approx 1$, $y_{23}$ opens the $c\tau$ channel while $y_{32}$ opens the $t\mu$ channel. This enlarges the Leptoquark width, diminishing $BR(j\mu)$ while maintaining $\sum_{i}BR(j\nu_{i})=1/2$ because of the neutrino mixing matrix. As the final states do not target top quarks nor tau leptons, this increases still more the difference between $\mu\mu jj$ and $\mu\nu jj$ final states in pair production. Single and non-resonant production remains relatively unchanged if one does not alter the values of $y_{22}$ and $y_{33}$.

In Section \ref{section:2} we decided to work with the $Q=1/3$ Leptoquark, although also a $Q=2/3$ Leptoquark could yield the same final states.  The reason for this decision is clear after the discussion in Section \ref{pheno}.  A Leptoquark with charge $Q=2/3$ would require more up-type instead of down-type quarks in the initial state to enhance the $\mu\nu jj$ final state, since this Leptoquark connects the neutrinos with $Q=2/3$ quarks.  Whereas in addition the third generation decay would be to $\tau \bar b$, which is not exactly $\nu j$ and acceptances analyses should be performed after the $\tau$ decay to investigate in which specific cases one could obtain an enhancement of $\mu\nu jj$ over $\mu\mu jj$. Therefore, the reasonable decision is to assume a Leptoquark with electric charge $Q=1/3$ in which case all the wanted features come out naturally.

Along this article we have focused in the excess reported by the CMS paper Ref.~\cite{Sirunyan:2018ryt} in the $\mu\nu j j$ final state. However, the same paper reports a deficit in the $\mu\mu jj$ final state at lower signal regions $M_{LQ}\approx 600\mbox{ GeV }\sim 800$ GeV. If one should try to interpret this with Leptoquarks, single- and non-resonant production come in handy as they can interfere with the SM. This interference is larger for smaller values in $M_{LQ}$. We computed with {\tt Madgraph\;5}~\cite{Alwall:2014hca} the interference between NP and SM main backgrounds $Z$+jets and found that in the benchmark point detailed in Eq.~\ref{eq:950} the interference is always destructive in $\mu\mu jj$. However, the strength of this interference is negligible and cannot account for the deviation. We have also investigated if another Leptoquark of mass $\sim 600$ GeV could be producing such an effect. We find in general that the interference in the $\mu\mu jj$ final state is destructive, but is not possible to reach the observed strength in the data while keeping the model from being ruled out by other direct Leptoquark searches. It is interesting to notice that, on the other hand, the $\mu\nu jj$ final state interferes with the SM background $W$+jets, but its sign can be adjusted by the relative sign between the PMNS component $U_{22}$ and the CKM matrix. As a curiosity, if one compares $S_{3}^{1/3}$ and $S_{1}^{1/3}$ as separate candidates with the same coupling constants, the interference in $\mu\nu jj$ is of opposite sign. This could be potentially useful to distinguish models.

\section{Conclusions}
\label{section:6}
Along this article we have considered single-, non-resonant, and pair production effects from a Leptoquark \change{with diagonal} couplings to second and third generation in the final states $\mu\mu jj$ and $\mu\nu jj$. We have shown that a non-resonant excess in the $\mu\nu jj$ final state and a fainter excess in the $\mu\mu jj$ final state is a pattern of this NP if the couplings are large enough and couple more the third than the second generation.

We have presented a simple Leptoquark model with ${\cal O}(1)$ couplings to second and third generation and a mass of ${\cal O}(1 \mbox{ TeV})$. On one hand, the strength of these couplings assures that single- and non-resonant effect are important and any kinematic resonant effect in the final states is washed out. On the other hand, a larger coupling to third generation favors the $\mu\nu jj$ final state by mainly two reasons: {\it i)} as long as there is no $b$-jet veto, since the $\nu$ is not flavor-tagged, a produced Leptoquark can decay to $j\nu$ where $j$ can be a $b$. {\it ii)} Single- and non-resonant production requires quarks in the initial state, and there is a larger abundance of the down-type quarks that produce the $\mu\nu jj$ final state than their same generation up-type quarks, whose corresponding diagrams would yield $\mu\mu jj$; with $\mu\nu jj$ having also $gc$ initial states. We have performed a simple simulation including the NP indicated by this model and compared our results to the recent CMS paper Ref.~\cite{Sirunyan:2018ryt}, where a non-resonant excess in the $\mu\nu jj$ has been reported. We have found that this simple Leptoquark model can easily accommodate the data better than SM alone and provide the features previously described. In particular, if we take the bin reported as the largest excess in $\mu\nu jj$ in CMS paper Ref.~\cite{Sirunyan:2018ryt}, then our model reduces the significance from $2.25\sigma$ to $0.75\sigma$ while the $\mu\mu jj$ final state is barely affected.

This work indicates that an excess of this kind, if produced by NP, would not be properly distinguished by current second generation Leptoquark searches at the LHC. The analysis in this article points out that to observe a NP signal of this kind, experimentalists should test $b$-tag on the jets. In this case the $\mu\nu jj$ sample will enhance its excess for the $b$-tagged events, whereas the excess should diminish in the $b$-tag veto sample. Moreover, if the statistic is large enough then pairing the $b$-jet with the $\nu$ and the light jet with the $\mu$ in the $\mu\nu jj$ sample could enhance a possible peak depending on the absolute value of the couplings. In addition to all this, it would be important for the present work if ATLAS would include the $\mu\nu jj$ final state in their analyses.

We have also discussed general features of the NP model. We find that the parameter space of the model has room for variations and still reproduce the same qualitative results. \change{We have found that the large couplings required to yield non-resonant effects would enter into conflict with low-energy observables.  We discuss in the Appendix how adding a new heavier Leptoquark could avoid low-energy flavor problems, and present the details of a model where this is accomplished.}  We also discussed another feature of the CMS paper Ref.~\cite{Sirunyan:2018ryt} that consists in a deficit in the $\mu\mu jj$ final state with respect to the SM expected background in the signal regions $M_{LQ} \approx 600$ GeV $\sim 800$ GeV. We show that SM and NP interference has negative sign in this final state. However we cannot fulfill the observed deficit within this simple Leptoquark model while keeping the model from being ruled out by other direct searches. It would be interesting to further investigate these or others Leptoquark models to also accommodate the deficit in $\mu\mu jj$, probably with more light Leptoquarks without altering present results.

Summarizing, we have shown that current second generation Leptoquark searches in the $\mu\mu jj$ and $\mu\nu jj$ final states are also sensitive to \change{diagonal} couplings to third generation as long as $b$-jet veto is not applied. Nevertheless, a $b$-tag handle can be useful to differentiate potential NP. We have presented a simple model that reproduces an excess recently found in CMS and that was disregarded because only second generation couplings were taken into account. 

This article suggests that Leptoquark searches at the LHC should take into account non-resonant effects and multigeneration couplings.

\section*{Acknowledgments}

We thank Leandro Da Rold, Aurelio Juste, David Morse and Olcyr Sumensari for very useful conversations in the subject of this work.

\appendix
\section{Flavour constraints overview}
\label{appendix}
\setcounter{equation}{0}

\renewcommand{\theequation}{\thesection.\arabic{equation}}

Along this Appendix we examine constraints coming from low-energy flavor physics that could enter into conflict with the large couplings required in the model presented in the article.  We find that such large couplings of ${\cal O}(1)$ would be ruled out by flavor constraints.  This could be avoided if other heavier particles are added to the model to cancel low-energy bounds while slightly affecting collider observables.  As a matter of fact, since the low-energy contributions go in general as (coupling/mass)$^2$, whereas contributions to collider physics observables have an exponential suppression in the mass due to the energy availability in the PDFs, this kind of cancellation is in general possible.

Since at this stage a new particle would be added ad-hoc to cancel the low-energy constraints, and in order to emphasize that we are presenting a proof of concept for the second generation Leptoquark searches, we explicitly present this alteration of the original model in the Appendix.  The objective of this Appendix is to show that is possible and with a variety of solutions to have the required cancellations at low-energy while having the non-resonant observed excess in $\mu\nu jj$.

We take as a departure point the model presented in Section \ref{section:2}.  We assume that the running of the couplings from the TeV scale to the low-energy due to the renormalization group equations is sub-leading as is usually done \cite{Gonzalez-Alonso:2017iyc, Feruglio:2018fxo}.  This running can be computed as in Ref.~\cite{Feruglio:2017rjo, Gonzalez-Alonso:2017iyc, Feruglio:2018fxo} and is expected to be a small contribution.  In any case, in the following paragraphs we show that the cancellations that take place due to adding a new particle have still a freedom of choice that could cancel as well this kind of contributions.

We find that the most severe constraint to the model from low-energy observables comes from the sensitive decay $B \to K^{(*)}\nu\nu$ \cite{Becirevic:2018afm, Marzocca:2018wcf} which was constrained to be \cite{Grygier:2017tzo} 
\begin{eqnarray}
R^{(*)}_{\nu\nu}&=&\frac{BR(B \to K^{(*)}\nu\nu)}{BR(B \to K^{(*)}\nu\nu)^{SM}} \nonumber\\
R_{\nu\nu} &<& 3.9 \nonumber\\
R^{*}_{\nu\nu} &<& 2.7 
\label{eq:knunu}
\end{eqnarray}
at 90$\%$ confidence level. The relevant Leptoquark effects can be parameterized as
\begin{eqnarray}
R^{*}_{\nu\nu}&=&\frac{\sum_{i,j}|\delta_{ij}C^{SM}_{L}+\delta C^{ij}_{L}|^2}{3|C^{SM}_{L}|^2} \nonumber \\
C^{SM}_{L} &=& -6.38(6) \nonumber\\
\delta C^{ij}_{L} &=& \frac{\pi v^2}{2 \alpha_{em}V_{tb}V^{*}_{ts}}\frac{y_{3j}y^{*}_{2i}}{M^2_{S^{1/3}}} 
\end{eqnarray}
Where the sum is over the neutrinos flavor. Considering our ansatz and the central values reported in Ref.~\cite{Tanabashi:2018oca} for every quantity, Eq.~\ref{eq:knunu} yields
\begin{equation}
\frac{|y_{33}y^{*}_{22}|}{M^2_{S^{1/3}}} < 0.045 \left(\frac{1}{\mbox{TeV}}\right)^2.
\label{flavorbound}
\end{equation}
Here we used the central value for $V_{ts}$ coming from indirect measurements assuming CKM unitarity \cite{Tanabashi:2018oca}.  However, for the sake of completeness, we also show in the following analysis the limits coming from direct bounds on $V_{ts}$ due to measurements of b-jet fractions in $t \to W j$ which is $V_{ts}<0.21$ at 95\% C.L.~\cite{Alvarez:2017ybk,Khachatryan:2014nda}; and does not assume CKM unitarity.

Since our original model ansatz in Section \ref{section:2} would in general not satisfy bounds in Eq.~\ref{flavorbound}, we add a second heavier $S^{'}_{1}$ with the same ansatz but different couplings $y^{'}_{22,33}$ and all others $y'_{ij}=0$.  This converts the bounds in Eq.~\ref{flavorbound} into
\begin{equation}
|\frac{y_{33}y^{*}_{22}}{M^2_{S^{1/3}}}+\frac{y^{'}_{33}y^{'*}_{22}}{M'^2_{S^{1/3}}}| < 0.045 \left(\frac{1}{\mbox{TeV}}\right)^2 .
\label{flavorbound2}
\end{equation}
With a correct relative sign assignment to the couplings it is possible to obtain a cancellation to the flavor contribution and satisfy Eq.~\ref{flavorbound2} while having non-resonant effects at the collider observables $\mu\mu jj$ and $\mu\nu jj$ as shown below.

Regarding other flavor constraints in this new model with a second heavier Leptoquark, charged currents $b \to c \ell \nu$ effects must be considered. LFU tests give hints on leptoquarks that couple to $\tau$-leptons while constraining couplings to $e$, $\mu$ leptons. The chosen ansatz does not produce scalar and tensor operators in the effective theory due to the absence of right-handed couplings; the Leptoquark contribution rescales the SMEFT \cite{Dorsner:2016wpm, Dorsner:2017ufx, Becirevic:2018afm, Marzocca:2018wcf}. For $B \to D^{(*)}_{s} \ell \nu$, the Leptoquark leading contribution is proportional to 

\begin{equation}
y_{3\ell}(y^{*}_{3\ell}+\frac{V_{cs}}{V_{cb}}y^{*}_{2\ell})
\end{equation}
It is easy to see that this vanishes for $\ell = e, \mu$ but not for $\tau$. Considering $R_{D^{*}}$, the Leptoquark effects at leading order are 
\begin{eqnarray}
\frac{R_{D}}{R^{SM}_{D}}&=&\frac{R_{D^*}}{R^{SM}_{D^*}} = 1.237 \pm 0.053 \approx 1 + 2g_{V_{L}} \nonumber \\
g_{V_{L}}&=&\frac{v^2}{4}\left(\frac{|y_{33}|^2}{M^2_{S^{1/3}}}+\frac{|y^{'}_{33}|^2}{M^{'2}_{S^{1/3}}}\right) .
\end{eqnarray}
Therefore $R_{D^{*}}$ implies 
\begin{equation}
2.466 < \sqrt{\left(|y_{33}|\frac{\mbox{TeV}}{M_{S^{1/3}}}\right)^2+\left(|y^{'}_{33}|\frac{\mbox{TeV}}{M^{'}_{S^{1/3}}}\right)^2} < 3.096
\label{RD}
\end{equation}
at the $1\sigma$ level.

Charged current effects are also seen in meson decays such as $B_{c} \to \ell \nu$ and $D_{s} \to \ell \nu$\cite{Dorsner:2016wpm, Dorsner:2017ufx}. The only relevant decays in the benchmark point (due to $y^{(')}_{23}=y^{(')}_{32}=0$) at leading order in $v/M^{(')}_{S^{1/3}}$ are
\begin{eqnarray}
\Gamma_{B_{c}\to \tau \nu}&=&\frac{G^{2}_{F}}{8\pi}f^2_{B_{c}}m^{3}_{B_{c}}(1-\frac{m^{2}_{\tau}}{m^{2}_{B_{c}}})^2\frac{m^{2}_{\tau}}{m^{2}_{B_{c}}}|V_{cb}|^2(1+\frac{v^2}{2M^{2}_{S^{1/3}}}|y_{33}|^{2}+\frac{v^2}{2M^{'2}_{S^{1/3}}}|y^{'}_{33}|^2)\nonumber\\
\Gamma_{D_{s}\to \mu \nu}&=&\frac{G^{2}_{F}}{8\pi}f^2_{D_{s}}m^{3}_{D_{s}}(1-\frac{m^{2}_{\mu}}{m^{2}_{D_{s}}})^2\frac{m^{2}_{\mu}}{m^{2}_{D_{s}}}|V_{cs}|^2(1+\frac{v^2}{2M^{2}_{S^{1/3}}}|y_{22}|^{2}+\frac{v^2}{2M^{'2}_{S^{1/3}}}|y^{'}_{22}|^{2})\nonumber
\end{eqnarray}
Constraints from these decays require form factors computed by lattice QCD\cite{Tanabashi:2018oca}. These expressions show that Leptoquark effect in these decays produce a rescaling of $V_{cb}$ and $V_{cs}$. Assuming that this scaling could be detected within the measured uncertainty in the CKM elements, the more relevant constraint would come from $V_{cs}$, which is measured with a percent-level precision \cite{Tanabashi:2018oca}.  Therefore,
\begin{equation}
|V_{cs}| \to |V_{cs}|(1+\frac{v^2}{4M^{2}_{S^{1/3}}}|y_{22}|^{2}+\frac{v^2}{4M^{'2}_{S^{1/3}}}|y^{'}_{22}|^{2})
\end{equation}
implies
\begin{equation}
\left(|y_{22}|\frac{\mbox{TeV}}{M_{S^{1/3}}}\right)^2+\left(|y^{'}_{22}|\frac{\mbox{TeV}}{M^{'}_{S^{1/3}}}\right)^2 < 1.13 
\label{Vcs}
\end{equation}
at the $1\sigma$ level.

\begin{figure}[h!]
\begin{center}
{\includegraphics[width=1\textwidth]{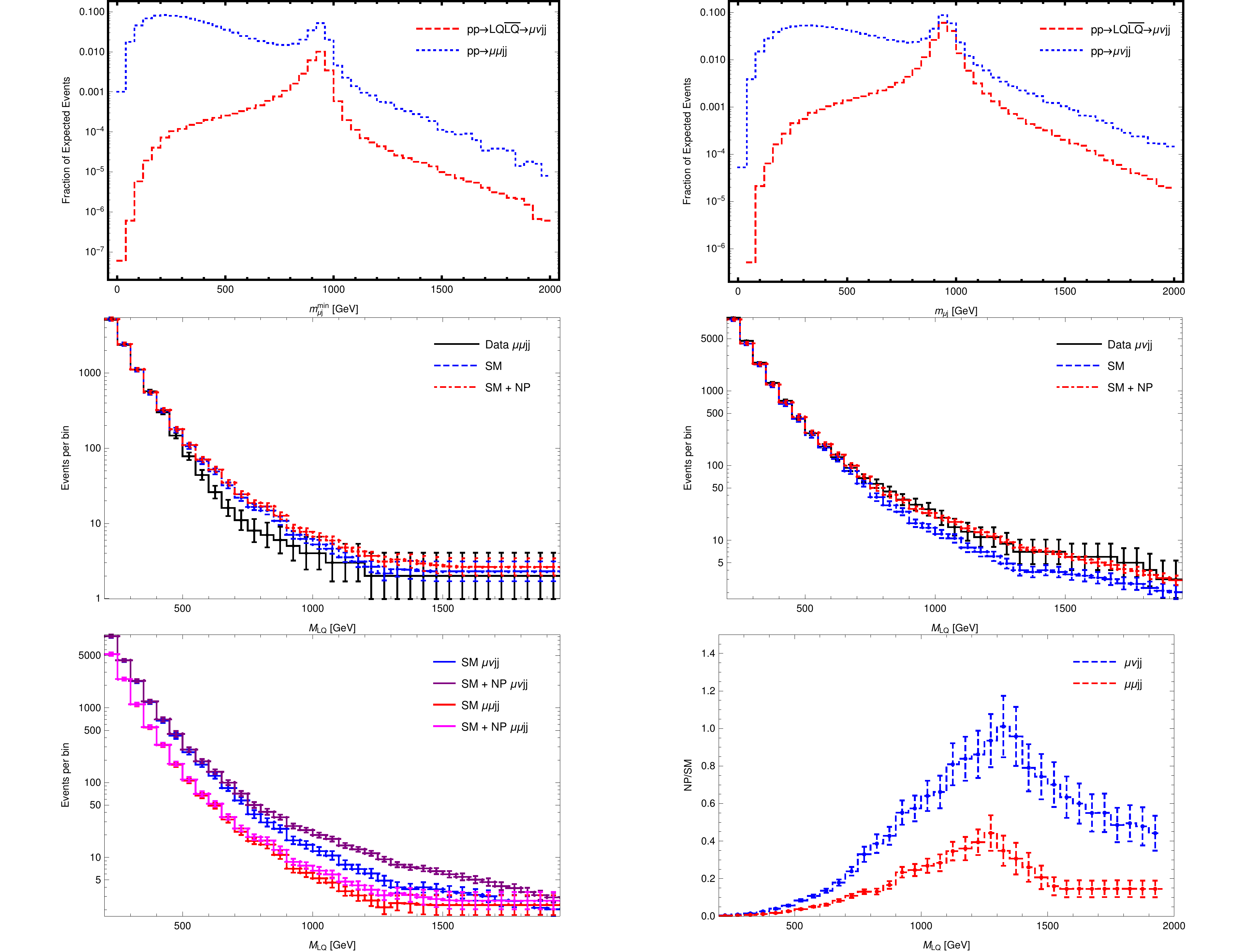}}
	\caption{Collider phenomenology of two $S_{1}$ in the benchmark point $(M_{S^{1/3}},\,y_{22},\,y_{33}) = (950 \mbox{ GeV},\,0.7,\, 1.2)$ and $(M'_{S^{1/3}},\,y'_{22},\,y'_{33}) = (1500 \mbox{ GeV},\,0.7,\, 3)$.  This model avoids the discussed flavor constraints, and it has still freedom in the parameters of the heavier Leptoquark as shown in Figs.~\ref{apendice2} and \ref{apendice3}.}
\label{apendice1}
\end{center}
\end{figure}

Having expressed quantitatively the constraints coming from low-energy precision physics, we proceed to find possible solutions that can provide non-resonant effects in collider observables while being safe to these constraints.  In order to explicitly construct a model, we choose a new benchmark point $\mbox{BP}^{'}$ where one Leptoquark has mass $M_{S^{1/3}}=950$ GeV and the second Leptoquark has mass $M'_{S^{1/3}}=1500$ GeV.  The only non-zero couplings are $(y_{22},\,y_{33}) = (0.7,\, 1.2)$ and $(y'_{22},\,y'_{33}) = (0.7,\, 3)$, which yields an exact cancellation in Eq.~\ref{flavorbound2} while still having couplings below the perturbative limit.  We have performed the same simulation as in Section \ref{section:4} and obtained the results shown in Fig.~\ref{apendice1}.  As it can be seen, this model with two Leptoquarks again reproduces the sought phenomenology regarding CMS paper \cite{Sirunyan:2018ryt} and, as a matter of fact, the presence of the heavy Leptoquark is hardly seen in the phenomenology.  The excess in the $M_{LQ}=950$ GeV bin in $\mu\nu jj$ is reduced in this case from 2.25 to 0.65, in contrast to 0.75 for the case of only one Leptoquark.

\begin{figure}[h!]
\begin{center}
\includegraphics[width=0.45\textwidth]{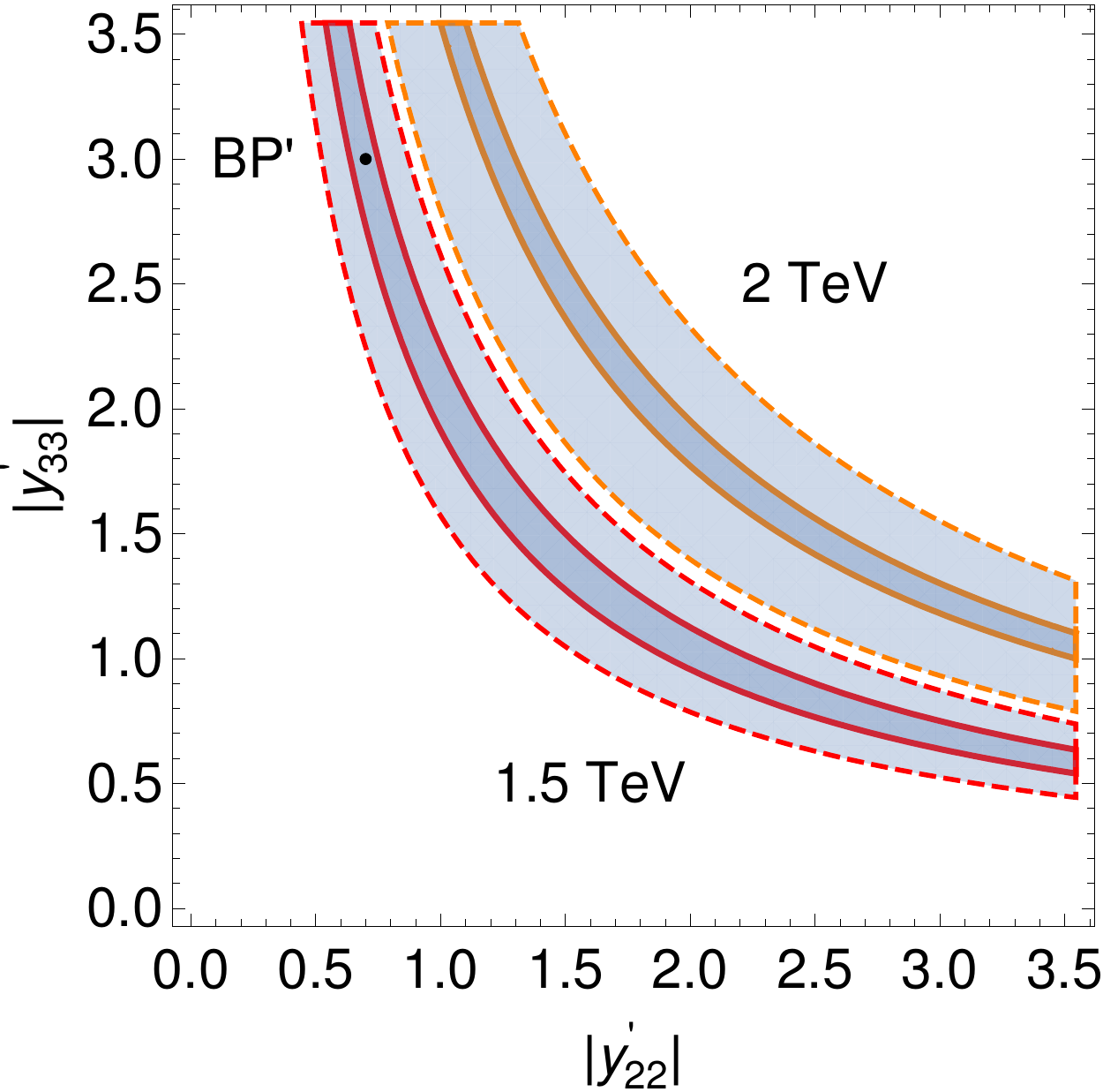}\hspace{3mm}
\includegraphics[width=0.45\textwidth]{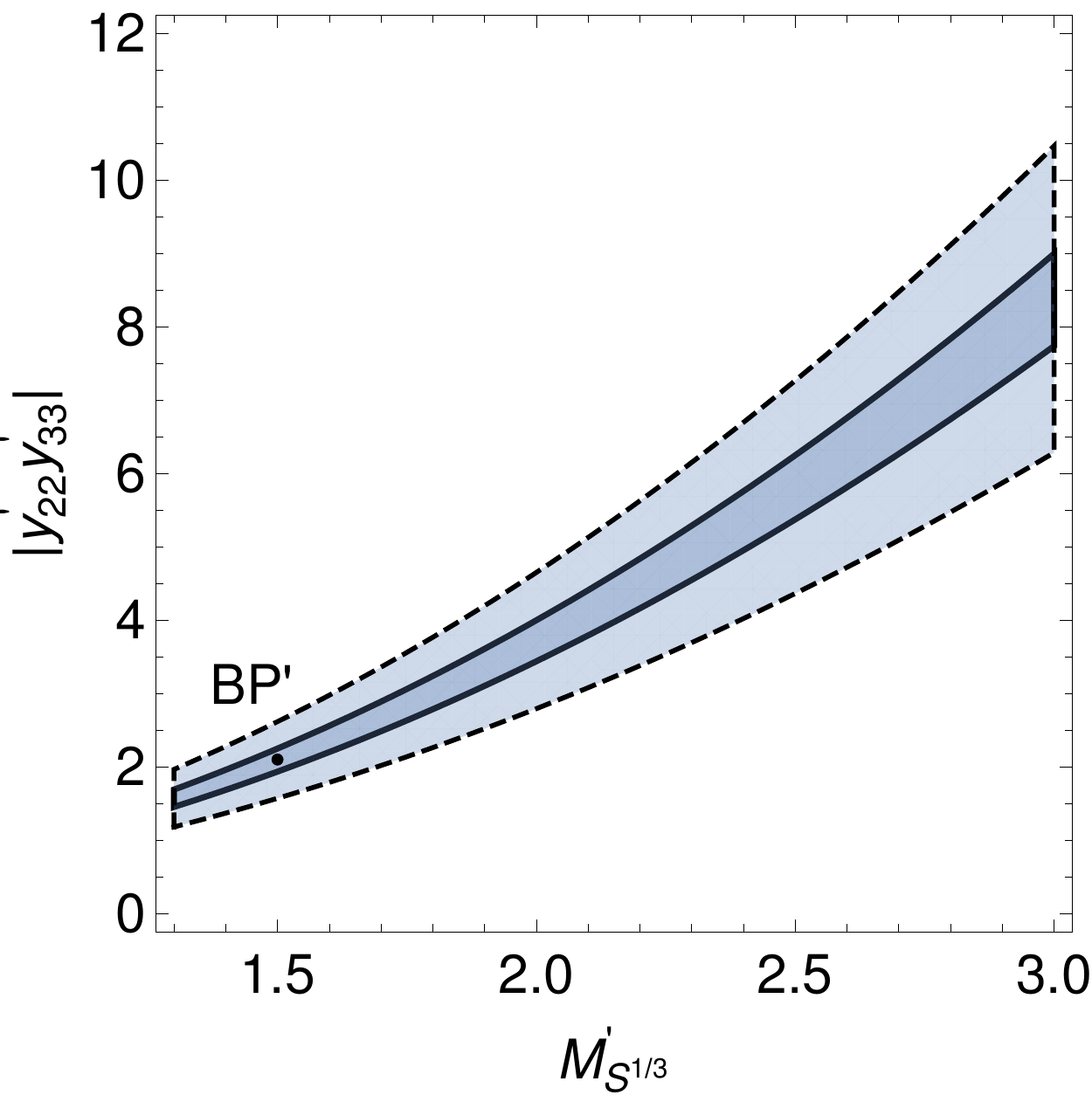}
	\caption{Allowed region in parameter space of the heavier Leptoquark where the $K^{(*)}\nu\nu$ constraint is satisfied once the lighter Leptoquark is fixed to $(M_{S^{1/3}},\,y_{22},\,y_{33}) = (950 \mbox{ GeV},\,0.7,\, 1.2)$. (Eq.~\ref{flavorbound2}).  In the left panel we consider two possible masses for the heavier Leptoquark.  The solid lines represent the 90\% C.L.~allowed using the central value for $V_{ts}$ assuming CKM unitarity, whereas the dashed lines correspond to the direct limit on $V_{ts}$ as detailed in text.}
\label{apendice2}
\end{center}
\end{figure}

\begin{figure}[h!]
\begin{center}
\includegraphics[width=0.45\textwidth]{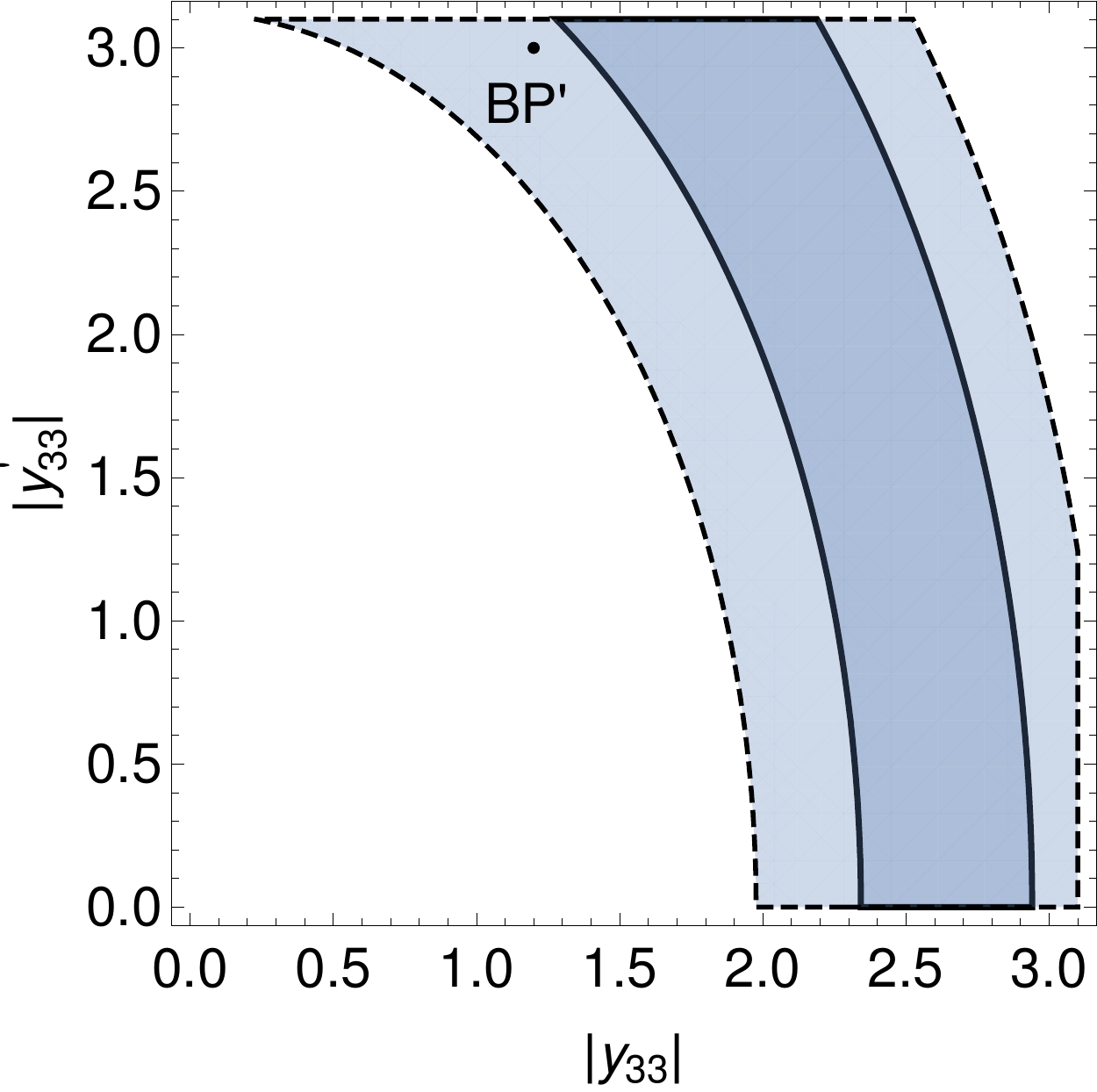}\hspace{3mm}
\includegraphics[width=0.45\textwidth]{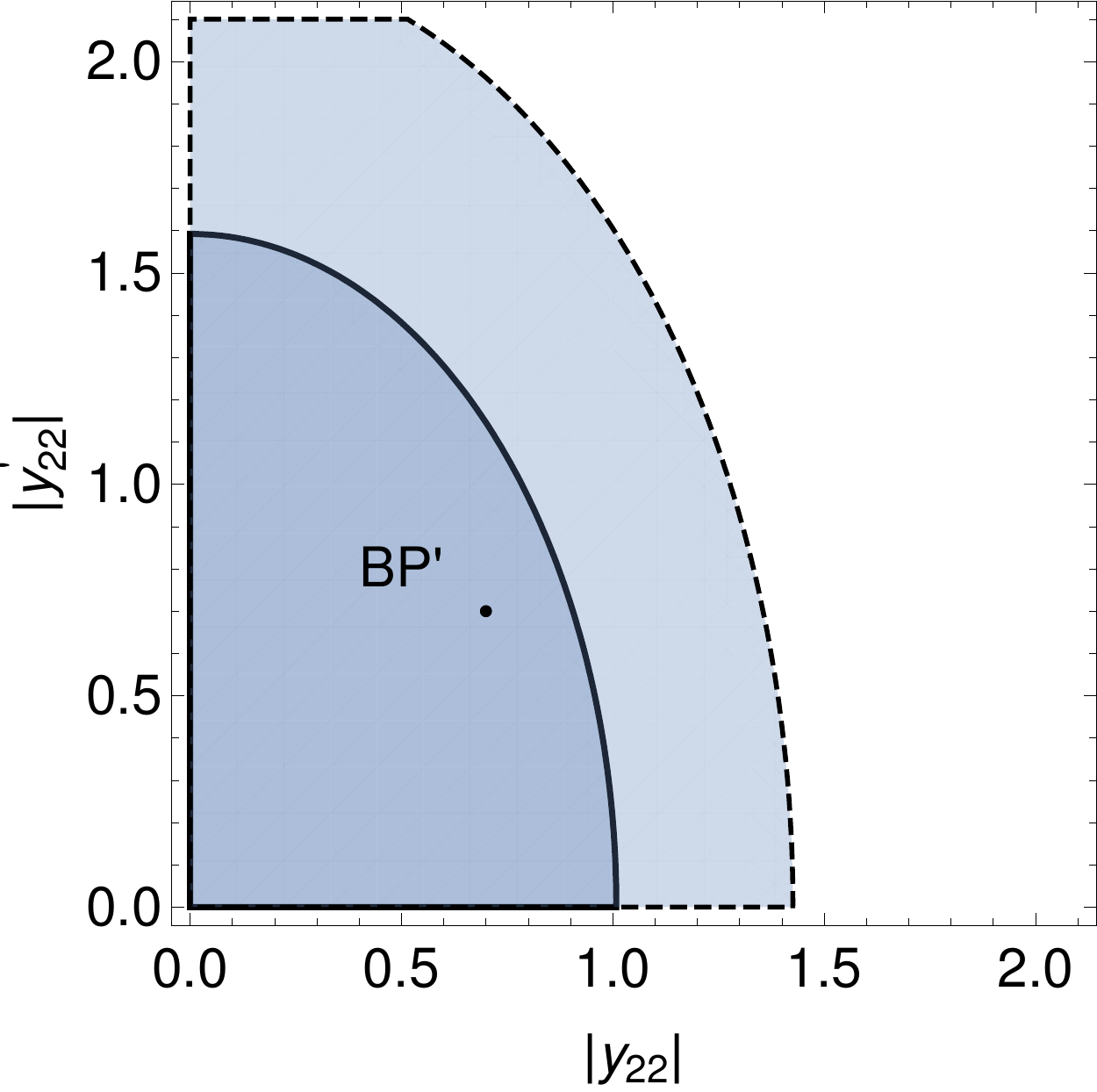}
	\caption{{\it Left:} Allowed region in parameter space of the heavier Leptoquark where the $R_D^{(*)}$ constraint is satisfied for $M'_{S^{1/3}}=1.5$ TeV and $(M_{S^{1/3}},\,y_{22},\,y_{33}) = (950 \mbox{ GeV},\,0.7,\, 1.2)$ (Eq.~\ref{RD}).  {\it Right:} idem for constraints coming from $V_{cs}$ precision (Eq.~\ref{Vcs}). The solid and dashed lines represent the $1\sigma$ and $2\sigma$ levels, respectively.  In both cases the mark corresponds to the selected benchmark point.}
\label{apendice3}
\end{center}
\end{figure}

Since the main collider phenomenology is produced by the 950 GeV Leptoquark, we can examine up to which extent the parameters of the heavier Leptoquark are determined by the flavor constraints.  In Fig.~\ref{apendice2} we study which is the freedom in the heavier Leptoquark parameters to still satisfy the $K^{(*)}\nu\nu$ constraint. In Fig.~\ref{apendice3} we analyze the freedom in parameter space for the heavier Leptoquark regarding constraints coming from $R^{(*)}_D$ and $V_{cs}$. 

Summarizing, we have verified that it is possible to have non-resonant effects in $\mu\mu j j$ and $\mu\nu jj$, with an excess in the latter, while avoiding the discussed flavor constraints.  Of course our model has too many arbitrary features, but it could be worth to study up to what extent these features could be obtained from a more complete theory.

\bibliographystyle{JHEP}
\bibliography{biblio}
\end{document}